\documentclass[journal]{IEEEtran}
\usepackage{makecell}
\usepackage{amssymb,amsthm,amsmath}
\usepackage{amsmath}
\usepackage{multirow}
\usepackage{algorithm}
\usepackage{algorithmicx}
\usepackage{algpseudocode}
\usepackage{lipsum}
\usepackage{graphicx}
\usepackage{multicol}
\usepackage{float}
\usepackage{url}
\usepackage{color}
\usepackage{stfloats}
\usepackage{booktabs}
\usepackage{bm}
\usepackage{graphicx}
\usepackage{rotating}
\usepackage{adjustbox}
\usepackage{verbatim}

\newcommand{\tabincell}[2]{\begin{tabular}{@{}#1@{}}#2\end{tabular}}

\algrenewcommand{\algorithmiccomment}[1]{\newline$\hfill$ #1}

\begin{document}
%
%\captionsetup[figure]{labelfont={bf},labelformat={default},labelsep=period,name={Fig.}}
%\captionsetup[figure]{name={Fig.},labelsep=period}
\title{GPU-Accelerated Compression and Visualization of Large-Scale Vessel Trajectories in Maritime IoT Industries}

\author{
	Yu Huang,~\IEEEmembership{Student Member,~IEEE,}
	Yan Li,~\IEEEmembership{Student Member,~IEEE,}
	Zhaofeng Zhang,\\
	and Ryan Wen Liu,~\IEEEmembership{Member,~IEEE}
	%
	%\thanks{\emph{Asterisk indicates the corresponding author.}}
	\thanks{The work was supported by the National Natural Science Foundation of China (No.: 51609195), and the Excellent Dissertation Cultivation Funds of Wuhan University of Technology (No.: 2018-YS-068). \textit{Yu Huang and Yan Li are joint first authors}. \textit{Corresponding authors: Zhaofeng Zhang and Ryan Wen Liu}.}
	\thanks{Y. Huang and Z. Zhang are with the Shanghai Advanced Research Institute, Chinese Academy of Sciences, School of Information Science and Technology, ShanghaiTech University, Shanghai 201210, China, and also with the University of Chinese Academy of Sciences, Beijing 100039, China (e-mail: huangyu@shanghaitech.edu.cn; zhangzf@sari.ac.cn).}
	\thanks{Y. Li and R. W. Liu are with the Hubei Key Lab. of Inland Shipping Technology, School of Navigation, Wuhan University of Technology, Wuhan 430063, China, and also with the Hubei Key Lab. of Transportation Internet of Things, School of Computer Science and Technology, Wuhan University of Technology, Wuhan 430070, China (e-mail: \{li\_yan, wenliu\}@whut.edu.cn).}
}
%
%
%\markboth{IEEE Internet of Things Journal}%
\markboth{Journal of \LaTeX\ Class Files,~Vol.~14, No.~8, August~2015}%
{Huang \emph{et al}.: GPU-Accelerated Compression and Visualization}
\maketitle

\begin{abstract}
	The automatic identification system (AIS), an automatic vessel-tracking system, has been widely adopted to perform intelligent traffic management and collision avoidance services in maritime Internet of Things (IoT) industries. With the rapid development of maritime transportation, tremendous numbers of AIS-based vessel trajectory data have been collected, which make trajectory data compression imperative and challenging. This paper mainly focuses on the compression and visualization of large-scale vessel trajectories and their Graphics Processing Unit (GPU)-accelerated implementations. The visualization was implemented to investigate the influence of compression on vessel trajectory data quality. In particular, the Douglas-Peucker (DP) and Kernel Density Estimation (KDE) algorithms, respectively utilized for trajectory compression and visualization, were significantly accelerated through the massively parallel computation capabilities of GPU architecture. Comprehensive experiments on trajectory compression and visualization have been conducted on large-scale AIS data of recording ship movements collected from $3$ different water areas, i.e., the South Channel of Yangtze River Estuary, the Chengshan Jiao Promontory, and the Zhoushan Islands. Experimental results illustrated that (1) the proposed GPU-based parallel implementation frameworks could significantly reduce the computational time for both trajectory compression and visualization; (2) the influence of compressed vessel trajectories on trajectory visualization could be negligible if the compression threshold was selected suitably; (3) the Gaussian kernel was capable of generating more appropriate KDE-based visualization performance by comparing with other seven kernel functions.
\end{abstract}

\IEEEpeerreviewmaketitle

\section{Introduction}
%

\begin{figure}[t]
	\centering
	\includegraphics[width=1.0\linewidth]{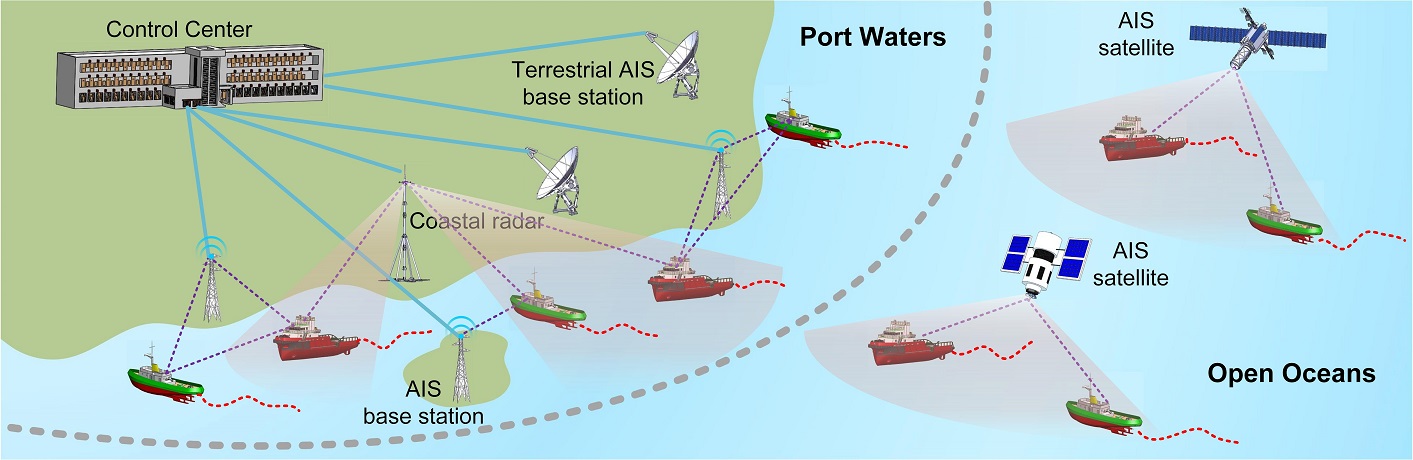}
	\caption{Overview of terrestrial and satellite AIS networks and their application scenarios in maritime IoT industries.}
	\label{Fig01_AISbackground}
\end{figure}

\IEEEPARstart{T}{he} automatic identification system (AIS), an automatic vessel-tracking system, has been widely adopted to identify and locate vessels through electronically exchanging both static and dynamic information with other nearby ships, terrestrial AIS base stations and satellites. It has the capacity of assisting in implementing intelligent traffic management and collision avoidance services in maritime Internet of Things (IoT) industries, illustrated in Fig. \ref{Fig01_AISbackground}. IoT is a new revolution of the Internet whose basic idea is the pervasive presence of a variety of things or objects \cite{RachediIEEEAccess2016, BendoudaFUTURE2018}. The International Maritime Organization (IMO) requires AIS transmitter to be fitted aboard all international voyage ships of $300$ gross tonnage and upwards and passenger ships \cite{YangWuTR2019}. As a consequence, AIS could provide a vast amount of near-real time information, i.e., spatio-temporal vessel trajectories, which  represent the moving behaviour of vessels. Vessel trajectory data mining techniques and their applications have recently gained increasing attention in both academia and industry \cite{LeiKIS2016,XiaoITS2017,TuITS2017,WangJoN2017,XiaoITS2019,LiuIoTJ2019}.
% and their applications in maritime IoT industries
%

To extract maritime traffic patterns behind massive AIS data, a variety of trajectory computing technologies \cite{FengZhuACCESS2016,ZhengTIST2015} have been developed to model the maritime vessel traffic. The traffic patterns are of importance in maritime situational awareness applications, e.g., traffic risk assessment, vessel behavior prediction, etc. Pallotta \textit{et al.} \cite{PallottaEntropy2013} proposed a AIS-based learning framework for traffic route extraction and vessel behavior modeling. Furthermore, the anomalous behavior detection \cite{LeiKIS2016,TerrosoISF2016,RiveiroDMKD2018} has recently attracted great attention due to its applications in maritime traffic surveillance and management. To automatically discover fishing areas, historical AIS data have contributed to the mapping of global (or local) fishing footprints \cite{KroodsmaScience2018,MazzarellaFUSION2014,BoerderSA2018,LeTixerantOCM2018} for enhancing fisheries management and fighting against illegal fishing activities in practice. Behind these successful activities, fundamental studies, such as trajectory reconstruction \cite{ZhangMengOE2018}, compression \cite{ZhangJoN2016} and clustering \cite{LiLiuACCESS2018}, etc., play important roles in making maritime applications more secure.
\subsection{Related Work}
This paper will pay more attention on the compression of massive vessel trajectories which could significantly reduce the computational cost required for maritime applications. The purpose of trajectory compression is to find a similar curve with fewer points to replace the original trajectory while preserving the important spatio-temporal geometrical properties \cite{HanTODS2017}. The rapid development of wireless multimedia sensor networks has made it easier to collect the spatio-temporal trajectory data \cite{BoulanouarADHOC2015, BoulanouarISCC2014}. In the era of big maritime data, tremendous numbers of vessel trajectories have been collected, which make maritime applications more time-consuming. To deal with the aforementioned limitations, Douglas-Peucker (DP) algorithm \cite{DouglasPeucker1973,MuckellGeoInformatica2014} and its extended versions \cite{ZhaoShiOE2018,ZhangShiJoN2019,LiuLiACCESS2019} have been developed to greatly reduce the storage requirements of massive vessel trajectories, leading to shortening computational time in different maritime IoT applications. To further improve compression results, both spatial and temporal information should be simultaneously adopted. A number of heuristic methods have been developed accordingly, such as Top-Down Time-Ratio (TD-TR) \cite{ZhangVLDB2018}, Spatial QUalIty Simplification Heuristic-Extended (SQUISH-E) \cite{MuckellGeoInformatica2014}, Threshold-Guided Algorithm \cite{PotamiasSSDBM06}, Generic Remote Trajectory Simplification (GRTS) \cite{LangePCC2009}, and Multiresolution Polygonal Approximation Algorithm \cite{ChenXuTIP2012}, etc. These heuristic methods, originally utilized for GPS trajectory compression, could be naturally extended to the compression of massive vessel trajectories. The implementations of these compression algorithms, however, easily suffer from high computational cost in the case of massive vessel trajectories. Fortunately, Graphics Processing Unit (GPU) \cite{OwensLuebkeCGF2007}, which has an excellent price-to-performance ratio, has emerged as an important parallel computing platform for computationally expensive problems. We will mainly consider GPU-accelerated DP algorithm since DP is the most popular trajectory compression algorithm. This acceleration framework could also be similarly extended to other trajectory compression algorithms.

Increasing attention has also been paid to the visualization of large data sets \cite{KeimTVCG2002}, which is able to assist in guaranteeing traffic safety through interactive visual exploration and analysis of massive movement trajectories. For example, the visualization of spatio-temporal trajectories is able to visually illustrate the spatio-temporal properties of trajectory data \cite{BuschmannVC2016}. Fishing is one of the most significant exploitative activities \cite{KroodsmaScience2018}. To further enhance the safety and security of fishing activities, the visualization frameworks for interactively visualizing fishing vessels have been developed in maritime IoT industries \cite{RakeshICCCNT2014,JamesMP2018}. More efforts have been achieved toward the visualization of vessel traffic density \cite{ScheepensTVCG2015,WuXuJoN2017}. It could help us understand and gain insights on vessel pattern knowledge from massive vessel trajectories. To detect the hotspot areas which have high potential to cause accidents, density map visualization methods have been developed to quantitatively depict the hotspot areas \cite{XiaoITS2017,ZhangTRPE2019}. The visualization methods have also been contributed to visually illustrating the vessel collision risk \cite{YooOE2018} and interactively detecting the vessels' abnormal behaviors \cite{RiveiroDMKD2018,ScheepensPVS2011}. For more recent studies on traffic visualization, please refer to the survey articles \cite{PackIEEECGA2010,ChenGuoITS2015}. In this work, the main focus is the visualization of vessel traffic density. It will also be utilized to investigate the influence of trajectory compression on density visualization quality. Several Kernel Density Estimation (KDE) algorithms, i.e., the density visualization means, will be implemented and compared under different visualization conditions.
\subsection{Motivation and Contributions}
It is well known that trajectory compression has become one of the most fundamental operations for trajectory data mining \cite{ZhengTIST2015}. However, the raw trajectory data is often extremely large \cite{VriesSomerenESA2012}, it will take high computational cost to perform trajectory compression through only CPU computing resources. Fortunately, multi-core CPUs and GPU \cite{LiJiangCG2013} have rapidly evolved into cost-effective computing frameworks, which have received considerable success in the fields of trajectory data mining, such as trajectory similarity search \cite{GowanlockTPDS2015,ZhangShenICDE2018}, trajectory clustering \cite{LohYuIS2015,GudmundssonValladaresTPDS2014}, trajectory classification \cite{ChangDekaICDM2012}, and trajectory prediction \cite{AltcheFortelleITSC2017}, etc. Without loss of generality, this paper will directly adopt the DP algorithm to compress the massive vessel trajectories to reduce the storage requirements. Due to the large amount of vessel trajectories in maritime IoT industries, it is necessary to shorten the trajectory compression cost while preserving the important topological features, which are associated with the vessel behaviors. Motivated by previous studies on GPU, we propose to tremendously accelerate trajectory compression by redesigning the hierarchical structure of traditional DP algorithm in the GPU computing platform.
\begin{figure}[t]
	\centering
	\includegraphics[width=1.0\linewidth]{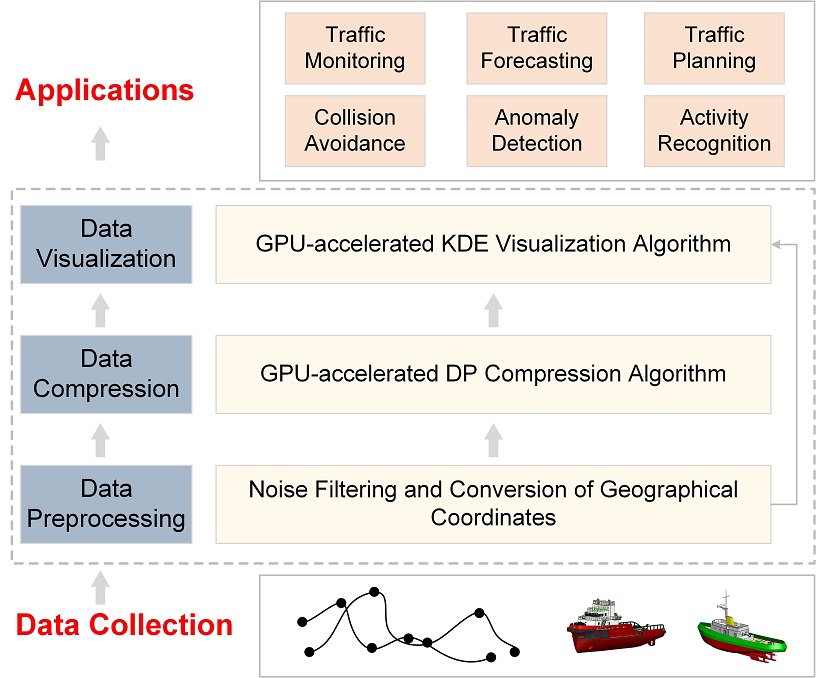}
	\caption{Flowchart of our GPU-based parallel implementation frameworks.}
	\label{Fig02_flowchart}
\end{figure}

To investigate the influence of compression on vessel trajectory data quality, GPU-accelerated visualization of vessel trajectories will also be implemented in this work. Data visualization has become an increasingly important visual analysis tool in traffic engineering \cite{ChenGuoITS2015,WangLiangICBDA2019,HeChenISPRS2019}. Due to the rapidly increasing amount of trajectory data, the low-efficiency trajectory visualization is now limiting the development of advanced technologies in practical applications. In the literature \cite{ZhangZhuGIS2017,HeimelKieferSIGMOD2015,BuschmannICC2014,BuergerTVCG2009}, GPU has emerged as a competitive parallel computing platform to significantly accelerate visualization of spatial-temporal data. KDE algorithm performs well in generating the density map visualization. The most popular KDE utilized in previous studies \cite{ScheepensTVCG2015,HurterErsoyCGF2012,PothkowHegeCGF2013} is the Gaussian kernel which generates smooth density estimate. To better understand the influences of different KDEs on visualization results, $8$ different kernels (i.e., Uniform, Triangular, Epanechnikov, Quartic, Triweight, Tricube, Gaussian and Cosine) will be parallelly implemented through the high performance computing of GPU.

To the best of our knowledge, no research has been implemented on GPU-accelerated DP algorithm for vessel trajectory compression. Compared with previous studies on accelerated visualization of massive data \cite{ZhangZhuGIS2017,HeimelKieferSIGMOD2015,BuschmannICC2014}, $8$ different KDE algorithms are parallelly performed in this work to assist in selecting the optimal one during trajectory data visualization. It is also worth noting that this is the first study focusing on GPU-based compression and visualization of massive vessel trajectories in maritime IoT industries. The flowchart of our proposed GPU-based parallel implementation framework is visually presented in Fig. \ref{Fig02_flowchart}. This study has a great significance in immensely reducing the computational time for both trajectory compression and visualization.

In conclusion, our main contributions, given the state-of-the-art research studies, can be summarized by the following three aspects:
\begin{enumerate}
	\item The hierarchical structure of DP-based trajectory compression algorithm has been redesigned according to GPU architecture and programming framework. It is able to significantly accelerate the compression of large-scale vessel trajectories while maintaining the compression quality.
	\item The GPU-based trajectory visualization framework has been proposed which mainly contains parallel data projection and interpolation, and memory communication reduction for accelerating KDE convolution. The visualization performance could be guaranteed while tremendously shortening the computational time under different experimental conditions.
	\item Experiments on large-scale vessel trajectories collected from $3$ different water areas have demonstrated that the proposed GPU-based parallelization frameworks consistently outperformed standard CPU computational baselines for both trajectory compression and visualization.
\end{enumerate}

The main benefit of our GPU-based parallelization frameworks is that it takes full advantage of the massively parallel computation capabilities of GPU architecture. The proposed acceleration frameworks thus have the capacity of enhancing the trajectory compression and visualization while dramatically shortening the computational time in maritime applications.
\subsection{Organization}
The remainder of this paper is organized as follows: Section \ref{sec:GPU} briefly reviews the recent advances in GPU. Section \ref{sec:methods} describes details on compression and visualization methods for AIS-based massive vessel trajectories. Section \ref{sec:GPUParallelization} is dedicated to the GPU-accelerated DP and KDE algorithms, respectively, utilized for compression and visualization of large-scale vessel trajectories. Comprehensive experiments on realistic vessel trajectories are carried out in Section \ref{section4}. This work is concluded by summarizing our main contributions in Section \ref{section5}.
% and discussing the future work
%
\begin{figure}[t]
	\centering
	\includegraphics[width=1.00\linewidth]{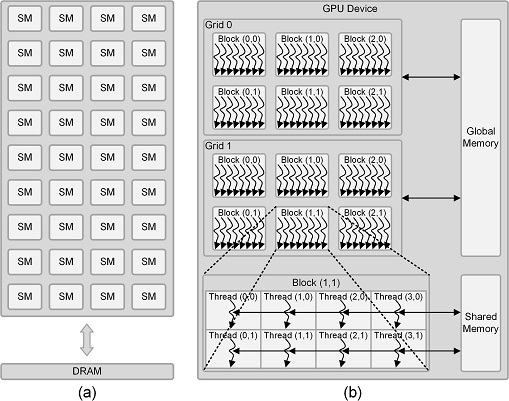}
	\caption{The schematic description of CUDA's architecture. From left to right: (a) NVIDIA GPU architecture, and (b) conceptual framework of CUDA programming model.}
	\label{Fig03_GPUframework}
\end{figure}
\section{Brief Review of GPU Computation}
\label{sec:GPU}
\begin{figure*}[t]
	\centering
	\includegraphics[width=0.9\linewidth]{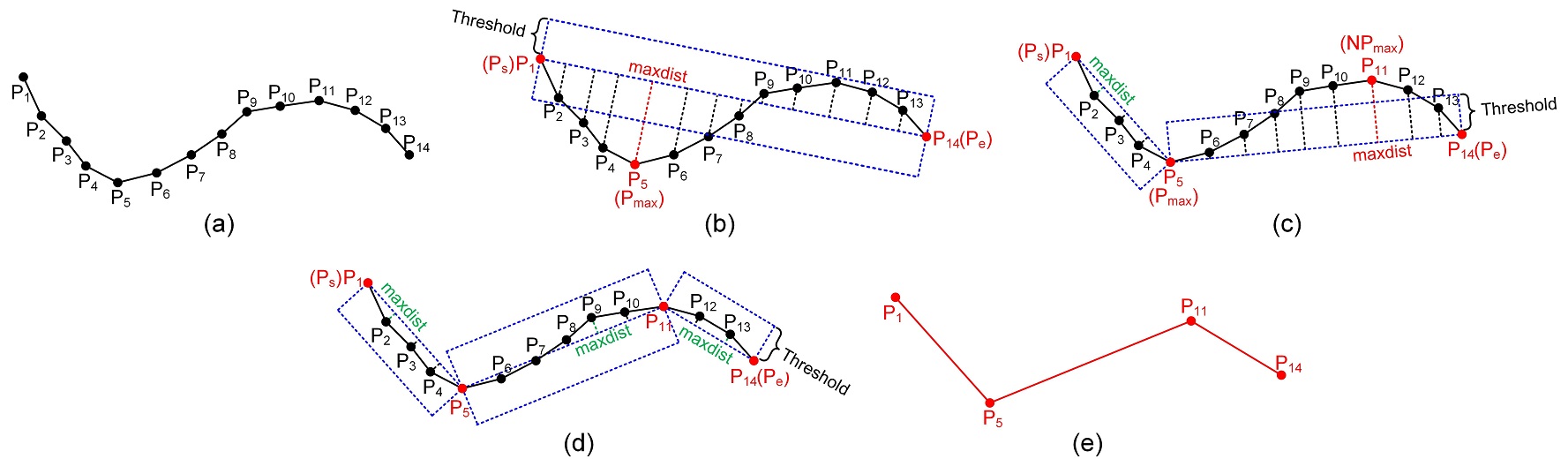}
	\caption{The schematic of basic Douglas-Peucker (DP) algorithm. From top-left to bottom-right: (a) original vessel trajectory, (b) searching feature point, (c) segmenting trajectory and searching new feature point, (d) iteratively searching important feature points, and (e) generating final compressed vessel trajectory.}
	\label{Fig04_DPProcessing}
\end{figure*}
Trajectory compression and visualization are essentially computationally demanding due to the large-scale AIS datasets. Traditional CPU serial implementations easily suffer from high computational cost in practical applications. It is fortunate that GPU, which can offer the excellent price-to-performance ratio, has become a competitive parallel computing platform for computationally expensive tasks in the field of big data mining \cite{ChenLiTPDS2018}. Current mainstream CPU commonly has a few cores, each of which simulates $2$ threads to perform computational operations. In contrast, the common-level GPU can concurrently execute thousands of threads in parallel \cite{NickollsDallyIEEE2010}.

GPU offers superior processing power and memory bandwidth than CPU, enabling it to perform higher efficiency in parallel computing. However, it is quite complicated to directly utilize GPU for general purpose or non-graphics computations. To make GPU computing easier, the general-purpose GPU (GPGPU), which performs non-specialized computations, has been proposed by mapping general purpose applications onto graphics hardware \cite{OwensLuebkeCGF2007}. In particular, the GPGPU computations are performed by implementing the non-graphics algorithms on existing GPU hardware. Both GPU and GPGPU development platforms have attracted tremendous attentions, however, only the professional researchers familiar with graphics APIs could smoothly utilize them in practical applications. It thus causes inconveniences for unfamiliar developers to make high-performance parallel computations available. To overcome these shortages, NVIDIA Corporation introduced the pioneering Compute Unified Device Architecture (CUDA) in late 2006. It is a parallel computing platform and flexible programming interface to perform general-purpose computing tasks on NVIDIA GPUs \cite{CUDA2019}. The CUDA can provide the C-like development environment for developers to reduce the complexity of parallel programming.

This work will implement the compression and visualization of large-scale vessel trajectories using CUDA programming language on NVIDIA GPUs. The schematic description of CUDA's architecture utilized in this paper is visually illustrated in Fig. \ref{Fig03_GPUframework}. In particular, the NVIDIA GPU architecture consists of a fully programmable scalable array of stream multiprocessors (SMs), each of which schedules multiple thread-blocks in parallel, shown in Fig. \ref{Fig03_GPUframework}(a). SMs can transfer data through the global memory on GPU. From the programmers' perspective, CUDA model is a framework where thousands of threads are parallelly performed \cite{CUDA2019,BigData2018}. The SMs creates, manages, schedules and executes threads in groups of fixed-size parallel threads called warps. When a kernel function is invoked, its related grid is launched. The grid consists of blocks, and multiple blocks can be assigned to a SM. Their scheduling obeys a round-robin strategy, shown in Fig. \ref{Fig03_GPUframework}(b). After being assigned to a certain SM, the block is divided into multiple warps. All threads are executed by a single instruction in the single-instruction multiple-thread (SIMT) architecture. Each thread executes the same instruction, but possibly on different data. The numbers of blocks and threads can be flexibly configured by programmers.

Data parallel programming model is able to map the data elements into the parallel processing threads. Many applications that handle large-scale data sets can utilize the data parallel programming model to speed up computational tasks. The vessel trajectories can be discretely represented and the timestamped points of each trajectory are logically independent. It is thus suitable to leverage GPU programming model to increase the computational efficiency in maritime IoT industries.
\section{Compression and Visualization Methods for Large-Scale Vessel Trajectories}
\label{sec:methods}
This section will first utilize the Mercator projection to convert the geographic coordinates of timestamped points in vessel trajectories into Cartesian coordinates. Both DP and KDE algorithms are then introduced, respectively, to perform compression and visualization of large-scale vessel trajectories in maritime IoT industries.
\subsection{Conversion of Geographical Coordinates}
\label{subsec:conversion}
The calculation of spherical distance between two adjacent timestamped points in one vessel trajectory is complicated based on geographic coordinate system. It is thus impossible to directly utilize the original vessel trajectories to implement compression and visualization in maritime IoT industries. It is also computationally intractable to calculate the distance between a timestamped point and a line segment based on geographic coordinate system. Therefore, both DP and KDE algorithms should be performed on Cartesian coordinate system which can be induced by the Mercator projection. Let $\left( \lambda, \varphi \right)$ denote the longitude and latitude coordinates of a timestamped point in one vessel trajectory. For the timestamped point $\left( \lambda, \varphi \right)$, the corresponding Cartesian coordinates $\left( x, y \right)$ can be obtained through the Mercator projection as follows
\begin{equation}\label{expression1}
{r_{0} = \frac{a} {\sqrt{1-e^2 \sin^{2} \varphi_{0}}} \cos \varphi_{0}},
\end{equation}
\begin{equation}\label{expression2}
    q = \ln \tan \left( \frac{\pi}{4} + \frac{\varphi}{2} \right) + \frac{e}{2} \ln \frac{1 - e \sin \varphi}{1 + e \sin \varphi},
\end{equation}
\begin{equation}\label{expression3}
	{x = \lambda  r_{0}},
\end{equation}
\begin{equation}\label{expression4}
	{y = q  r_{0}},
\end{equation}
where $a$ is the long radius of earth ellipsoid, $\varphi_{0}$ is the standard latitude in Mercator projection, $r_{0}$ is the radius of parallel circle of standard latitude, $e$ is the first eccentricity of earth ellipsoid, and $q$ is the isometric latitude.
\subsection{DP-Based Vessel Trajectory Compression}
\label{originalDP}
The purpose of DP is to approximate the original vessel trajectory with a similar trajectory with fewer timestamped points (i.e., feature points). These preserved points are essentially related to the important vessel-specific behaviours under different navigation conditions. DP has gained tremendous attention due to its simplicity and efficiency in maritime applications. Its basic strategy is that the approximation must contain a subset of original timestamped points and all these points must lie within a preselected distance to the approximation. As shown in Fig. \ref{Fig04_DPProcessing}, let $T = \left\{ P_{1}, P_{2}, \cdots, P_{N_{o}} \right\}$ denote a vessel trajectory where $P_{n}$ with $n \in \left\{ 1, 2, \cdots, N_{o} \right\}$ represents the $n$-th timestamped point and $N_{o}$ is the length of vessel trajectory $T$. The procedure of DP is detailedly given as follows
\begin{enumerate}%[Step 1:]
	\item To preserve the important timestamped points, we first determine a pre-defined threshold $\epsilon$ as a criterion to compress the original trajectories. The first point ${P}_{1}$ and last point ${P}_{14}$ are respectively marked with $P_{s}$ and $P_{e}$, which are kept as the important feature points. The line $\overline{P_{s}P_{e}}$ connecting the first and last points is selected as the datum line (i.e., curve segment). The vertical Euclidean distance (VED) of each point to the datum line is serially calculated. The point (i.e., ${P}_{5}$) related to the maximum VED will be selected as the feature point and divides the original trajectory into two sub-trajectories, shown in Fig. \ref{Fig04_DPProcessing}(b).
	\item As illustrated in Figs. \ref{Fig04_DPProcessing}(c) and (d), it is still utilized to iteratively find the maximum VEDs among different sub-trajectories. The corresponding points (i.e., ${P}_{5}$ and ${P}_{11}$) are correspondingly selected as the feature points, which will be preserved in the compressed trajectory.
	\item DP algorithm recursively divides the original trajectory and find the timestamped points related to the maximum VEDs. This procedure will be implemented iteratively until there is no feature point whose VED is larger than the pre-defined threshold $\epsilon$. The final compressed trajectory is visually displayed in Fig. \ref{Fig04_DPProcessing}(e).
\end{enumerate}

The DP algorithm can be performed in ${O} (N_o \log N_o)$ time on average. Its worst-case time complexity is ${O} ( N_c N_o )$, where $N_o$ and $N_c$ denote the lengths of the original and compressed vessel trajectories, respectively \cite{ChenXuTIP2012}. It is thus an output-dependent compression algorithm and it will be very slow when the approximation is finer (i.e., $N_c$ is very large).
\subsection{KDE-Based Vessel Trajectory Visualization}
\label{sec:KDETraj}
This subsection mainly focuses on the visualization of large-scale vessel trajectories, which is essentially related to the vessel density visualization. Before the generation of visualization, it should first divide the water areas into a finite number of grids which contribute to the density matrix $\mathcal{M}_{D}$. Trajectory visualization will be performed on this density matrix with the size of  $u \times v$. Let  $(\tilde{x}, \tilde{y})$ with $1 \leq \tilde{x} \leq u$ and $1 \leq \tilde{y} \leq v$ be an element in $\mathcal{M}_{D}$. From a statistical point of view, the intensity of each element in $\mathcal{M}_{D}$ reflects the vessel density. To calculate the intensity, it is necessary to project the timestamped points in each trajectory into the corresponding elements in $\mathcal{M}_{D}$. As described in Section \ref{subsec:conversion}, geometric positions of timestamped points can be represented by the Cartesian coordinates $( x, y )$ through the Mercator projection. The conversion relationship between $(\tilde{x}, \tilde{y})$ in density matrix and $( x, y )$ in Cartesian coordinates is thus given as follows
\begin{equation}\label{expression5}
    \tilde{x}_{k}^{n} = \lceil \frac{x_{k}^{n} - x_{\min} }{ x_{\max} - x_{\min} }  \cdot \left( u - 1 \right) \rceil + 1 \in \left[ 1, u \right],
\end{equation}
with $x_{\min} = \min\limits_{1 \leq k \leq K, 1 \leq n \leq N} x_{k}^{n}$, $x_{\max} = \max\limits_{1 \leq k \leq K, 1 \leq n \leq N} x_{k}^{n}$, and
\begin{equation}\label{expression6}
    \tilde{y}_{k}^{n} = \lceil \frac{y_{k}^{n} - y_{\min} }{ y_{\max} - y_{\min} }  \cdot \left(v - 1 \right) \rceil + 1 \in \left[ 1, v \right],
\end{equation}
with $y_{\min} = \min\limits_{1 \leq k \leq K, 1 \leq n \leq N} y_{k}^{n}$, $y_{\max} = \max\limits_{1 \leq k \leq K, 1 \leq n \leq N} y_{k}^{n}$. Here, $K$ is the total number of vessel trajectories, $N$ is the total length of one vessel trajectory, $( x_{k}^{n}, y_{k}^{n} )$ denotes the Cartesian coordinate of the $n$-th timestamped point in the $k$-th vessel trajectory, and $\lceil \cdot \rceil$ denotes the ceil operation.

If the total number of timestamped points, whose projected coordinates are $(\tilde{x}, \tilde{y})$, is $C$, the intensity of element $(\tilde{x}, \tilde{y})$ in $\mathcal{M}_{D}$ becomes $\mathcal{M}_{D} (\tilde{x}, \tilde{y}) = C$ essentially indicating the density of vessel traffic. It is well known that vessel trajectories are commonly collected using AIS messages in discrete intervals of $2$-$180$ seconds \cite{WestrenenSMC2015}. The long interval easily gives rise to the loss of timestamped points in raw vessel trajectories, leading to low-quality visualization of vessel density. Theoretically, any vessel trajectory should contribute to a continuous path (i.e., continuous element) in $\mathcal{M}_{D}$. Two elements, yielded by two adjacent points in one raw trajectory, may easily be separated by several empty elements due to the long intervals. The existing interpolation algorithm can be directly adopted to estimate the missing location information between any two adjacent points if the AIS broadcast interval is long. More details will be discussed in Section \ref{sec:KDEparallelization}.
\begin{table}[t]
	\centering
	\caption{Eight different kernel functions utilized in GPU-accelerated trajectory visualization experiments.} \label{Table1}
	\renewcommand\arraystretch{2}
	\begin{tabular}{|c|l|}
		\hline
		Kernel Functions & Mathematical Formulas \\ \hline
		Uniform         & $f(s,t)= \left(\frac{1}{2}\right)^{2}I(s)I(t)$     \\ \hline
		Triangular      & $f(s,t)= \left(1 - |s|\right) \left(1 - |t|\right)I(s)I(t)$         \\ \hline
		Epanechnikov    & $f(s,t)= \left(\frac{3}{4}\right)^{2}\left(1-s^2\right)\left(1-t^2\right)I(s)I(t) $          \\ \hline
		Quartic         & $f(s,t)= \left(\frac{15}{16}\right)^{2}\left(1-s^2\right)^{2}\left(1-t^2\right)^{2}I(s)I(t) $          \\ \hline
		Triweight       & $f(s,t)= \left(\frac{35}{32}\right)^{2}\left(1- s^2\right)^{3}\left(1- t^2\right)^{3}I(s)I(t)$      \\ \hline
		Tricube         & $f(s,t)= \left(\frac{70}{81}\right)^{2}\left(1- |s|^3\right)^{3}\left(1- |t|^3\right)^{3}I(s)I(t)$          \\ \hline
		Gaussian        & $f(s,t)= \left(\frac{1}{\sqrt{2\pi}}\right)^{2}\exp \left( - \left(\frac{s^{2}}{2}+\frac{t^{2}}{2} \right) \right)$      \\ \hline
		Cosine          & $f(s,t)= \left(\frac{\pi}{4}\right)^{2} \cos \left( \frac{\pi}{2}s \right) \cos \left( \frac{\pi}{2}t \right) I(s)I(t) $                      \\ \hline
	\end{tabular}
\end{table}
\begin{figure*}[t]
	\centering
	\includegraphics[width=0.90\linewidth]{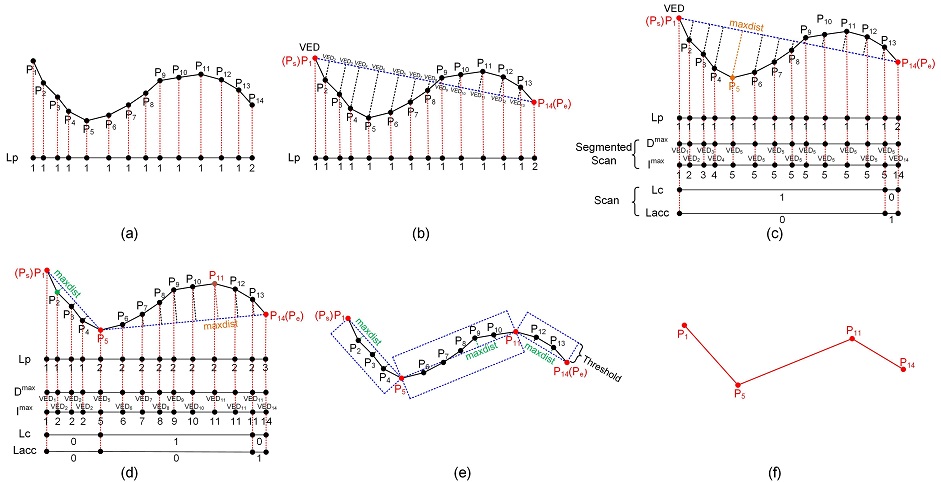}
	\caption{Procedures of our proposed GPU-based parallel compression of vessel trajectory. From top-left to bottom-right: (a) original vessel trajectory which is marked with label $Lp$, (b) parallel computation of VED, (c) respectively utilizing the segmented scan and scan algorithms to find the maximal VED and count the new feature points, (d) updating the searched feature points, (e) iteratively searching important feature points, and (f) generating final compressed vessel trajectory.}
	\label{Fig05_DPParallel}
\end{figure*}

From a data visualization point of view, the vessel traffic density map should be spatially smooth. If we directly utilize the estimated $\mathcal{M}_{D}$ as the density map, the visualization results will suffer from the non-smooth intensities. It is well known that KDE has become an important tool for visualizing the distribution of spatial data. To further improve the visualization performance, KDE will be utilized to generate a smoother density surface of $\mathcal{M}_{D}$ over the 2D geographic space. In particular, the final vessel traffic density map $\bar{\mathcal{M}}_{D}$ can be calculated as follows
\begin{equation}\label{expression7}
\begin{split}
\bar{\mathcal{M}}_{D} (\tilde{x}, \tilde{y}) &= {f} (s,t) \otimes \mathcal{M}_{D} (\tilde{x}, \tilde{y}) \\
            &= \sum\limits_{s = - a}^a {\sum\limits_{t =  - a}^a {f (s,t) \mathcal{M}_{D} } (\tilde{x} - s, \tilde{y} - t)},
\end{split}
\end{equation}
where $\otimes$ is a convolutional operator, $f(\cdot, \cdot)$ denotes the 2D kernel function of size $\varpi \times \varpi$. In the literatures, Gaussian kernel has become the most commonly used kernel function. A large variety of kernel functions have been further investigated, e.g., Uniform, Triangular, Epanechnikov, Quartic, Triweight, Tricube and Cosine, etc. Visualization experiments will be performed on these $8$ different kernel functions. The corresponding mathematical formulas are shown in Table \ref{Table1}. The definition of indicator function $I(\cdot)$ is given by
\begin{equation}\label{expression8}
I(w)=
\begin{cases}
1& \text{if $|w|\leq \frac{\varpi - 1}{2}$}, \\
0& \text{otherwise},
\end{cases}  \\
\end{equation}
where $\varpi$ denotes the kernel bandwidth which is a positive odd integer. It strongly effects the resulting vessel density surface. Experiments will be implemented in Section \ref{sec:expKDE} to determine the proper bandwidth of kernel function $f(\cdot, \cdot)$.
%
%where $\varpi \times \varpi$ denotes the size of convolution kernel.
%
%
\section{GPU-Based Parallelization Frameworks}
\label{sec:GPUParallelization}
In this section, we will give more details about the GPU-based parallelization frameworks for both DP compression and KDE visualization. According to the analysis of original DP algorithm, the basic computational strategy behind DP is the recursive calculations.Recursive programs usually have an associated recursion depth. For compression of vessel trajectories in this work, the recursion depth greatly increases as the compression threshold decreases. There is also a hardware limit on maximum recursion depth in GPU programming model, which may cause serious problem. It thus can not be directly applied in exist GPU-accelerated framework. To overcome this limitation, we propose to redesign the hierarchical structure of commonly-utilized DP algorithm. A point projection method is also introduced in the proposed parallel KDE framework to make visualization available to the large-scale vessel trajectories. Furthermore, the strategy of memory access is optimized to meet the requirements such as high speed and continuous data compression.
\subsection{Parallelization of DP}
\label{sec:PallaDP}
To maximize the degree of parallelization, we will redesign the hierarchical structure of DP to simultaneously process all of the curve segments generated in each iteration. To take full advantage of the GPU computing resources, the vessel trajectory will be divided into blocks by a thread grid. This mechanism can greatly improve the efficiency of trajectory compression. However, due to the characteristic of GPU programming model \cite{CUDA2019}, thread blocks are required to execute independently once the kernel function is being implemented. There is no communication among these thread blocks when they are running. This independence requirement allows thread blocks to be scheduled in any order across any number of cores \cite{CUDA2019}. Since timestamped points of trajectory compression must be ordered, it is necessary to leverage scan \cite{Scan2007} and segmented scan algorithms \cite{Fastscan2008} to solve the problem produced during parallel execution.
\subsubsection{Initialization}
Analogous to Section \ref{originalDP}, one vessel trajectory is represented by a series of timestamped points $T = \left\{P_{1},P_{2}, \cdots, P_{N_{o}} \right\}$. Let $T^* = \left\{ P^*_{1}, P^*_{2}, \cdots, P^*_{N_t} \right\}$ and $T' = \left\{ P'_{1}, P'_{2}, \cdots, P'_{N_c} \right\}$, respectively, denote the temporary and final compressed trajectory. It is easy to get $N_t \leq N_c \leq N_o$. As mentioned in Section \ref{originalDP}, multiple curve segments are generated to simplify original trajectory. At the beginning of compression, the whole trajectory is treated as one curve segment. The first and last points, extracted from the original trajectory $T$, are directly copied to $T^{'}$ while keeping $T^*$ empty.
% The Cartesian coordinate of each point $P_n$ in $T$ is $( x^{n}, y^{n} )$ with $1 \leq n \leq N_{o}$. 

To efficiently classify the timestamped points in different curve segments, we introduce a set of ordered labels $Lp = \left\{ L_{1}, L_{2}, \cdots, L_{N_{o}} \right\}$ for all points in the trajectory $T$ shown in Fig. \ref{Fig05_DPParallel}(a). Each point naturally corresponds to a unique label. The points related to the same curve segment thus have the same label value. By querying this label set, a certain points' curve segments can be quickly located. To make parallelization easier, we first let $L_{1 \leq n \leq N_{o}-1} = 1$ and $L_{N_{o}} = 2$ before iteration takes place.
\begin{algorithm}[t]
	\caption{Update Result Set} \label{alg:updateSet}
	\begin{algorithmic}[1]
		
		\Require $T^{'}$, $Is$, $Lacc$, $I^{\max}$,
		\Ensure $T^{*}$
		\State  int oldId = blockId * blockDim + threadId;
		\State  newId = oldId + $Lacc$[curveId];
		\Comment /* Step 1. Copy points from $T^{'}$ to $T^{*}$. */
		\If {($T^{'}$[oldId] != NULL)}
		$T^{*}$[newId] = $T^{'}$[oldId];
		\EndIf
		\Comment /* Step 2. Compute the new point index. */
		\State insertId = $I^{\max}$[$Is$[oldId + 1] - 1];
		\State newId++;
		\Comment /* Step3. Save new point to $T^{*}$. */
		\State $T^{*}$[newId] = $T$[insertId];
	\end{algorithmic}
\end{algorithm}
\subsubsection{Iteration}
GPU-based parallelization of the DP algorithm can be mainly divided into the following several steps:
\begin{enumerate}
	\item[a.] \textit{Compute VED for every point}. The set $T^{'}$ is essentially adopted to only store the start and end points of the generated curve segments. With the help of set $T$, the VED between each point and its related curve segment can be concurrently computed by threads shown in Fig. \ref{Fig05_DPParallel}(b). Given the $n$-th point $P_{n} \in T$ with $1 \leq n \leq N_o$, the VED $D_{n,i}$ between this point and the corresponding curve segment can be obtained as follows
	\begin{equation}\label{expression9}
	D_{n,i} = \frac{|\overrightarrow {P_{s,i}P_{n} } \times \overrightarrow {P_{s,i}P_{e,i}}|}{|\overrightarrow {P_{s,i}P_{e,i}}|},
	\end{equation}
	where $P_{s,i}$ and $P_{e,i}$, respectively, represent the start and end points of the $i$-th curve segment with $1 \leq i \leq I_{o}$. $I_{o}$ denotes the total number of curve segments generated in current iteration. In practice, $D_{n,i}$ is only calculated between $P_{n}$ and its related curve segment. This one-to-one relationship could be guaranteed through the introduced set $Lp$.
	\item[b.] \textit{Sort the maximal VED of every curve segment}. In Eq. (\ref{expression9}), the index $i$ is essentially specific for the $n$-th point $P_{n}$ due to the one-to-one relationship. For the sake of simplicity, we directly adopt $D_{n}$ instead of $D_{n,i}$ to denote the VED. As shown in Fig. \ref{Fig05_DPParallel}(c), we exploit the segmented scan algorithm to orderly scan the set $D = \left\{ D_{1}, D_{2}, \cdots, D_{N_{o}} \right\}$. The details on parallel implementation of segmented scan algorithm will be introduced in Section \ref{sec:SegScan}. With the help of segmented scan algorithm, we can obtain the timestamped point which has the maximal VED. Owing to the pre-defined set $Lp$, all curve segments can be scanned in parallel. Let $D^{\max} = \left\{ D^{\max}_{1}, D^{\max}_{2}, \cdots, D^{\max}_{N_{o}} \right\}$ and $I^{\max} = \left\{ I_{1}, I_{2}, \cdots, I_{N_{o}} \right\}$, respectively, represent the scanned results and their index values. For the sake of better understanding, we assume that $\left\{P_{n1}, P_{n1+1}, \cdots, P_{n2} \right\} \subseteq T$ with $1 \leq n1 < n2 \leq N_o$ has the same label value $L_{1 \leq i \leq I_{o}}$. Let $D_{n1 < n < n2}$ be related to the maximal VED, we can update $D^{\max}_{\bar{n}} = D_{\bar{n}}$ and $I_{\bar{n}} = \bar{n}$ if $n1 < \bar{n} < n$; while $D^{\max}_{\bar{n}} = D_{n}$ and $I_{\bar{n}} = n$ if $n \leq \bar{n} < n2$.
	\item[c.] \textit{Mark the points needed to retain}. If the maximal VED $D_n$ is larger than the pre-selected threshold $\epsilon$, $P_{n} \in T$ should be retained as a feature point. The label set $Lc = \{ L_{1}^{'} , L_{2}^{'} , \cdots, L_{I_{o}}^{'} \}$ is then introduced to mark the curve segment which corresponds to $D_n$. $Lc$ records the situation whether feature points are generated on the related curve segments. As shown in Fig. \ref{Fig05_DPParallel}(c), taking the feature point $P_{5}$ with the maximal VED as an example, we can set $L_{1}^{'} = 1$ and $L_{2}^{'} = 0$ with $I_{o} = 2$. The corresponding curve segment can then be divided into two parts. A new label set $Is = \{ I_{1}^{'}, I_{2}^{'}, \cdots, I_{I_{o}}^{'} \}$ is adopted to store the index of start point of the new curve segment, i.e.,
	\begin{equation}
	    I_{i}^{'} = n + 1,
	\end{equation}
	with $i > 1$. Here, $i$ denotes the $i$-th curve segment newly generated in the current iteration, and $n$ denotes the $n$-th timestamped point related to the maximal VED.
	\item[d.] \textit{Count the number of feature points}. The scan algorithm is utilized to scan the set $Lc$. The corresponding implementation details can be found in Section \ref{Sec:ScanSection}. There are two types of scan algorithms commonly utilized, i.e., inclusive scan and exclusive scan \cite{Scan2007}. The main difference between them is that result of the former contains the first element of input, but the latter one does not. In this paper, we propose to adopt the exclusive scan in our experiments. Let $Lacc = \{L_{1}^{*} , L_{2}^{*} , \cdots, L_{I_{o}}^{*} \}$ denote the scanned results. As shown in Fig. \ref{Fig05_DPParallel}(c), $L_{i}^{*}$ denotes the number of points needed to retain before yielding the $i$-th curve segment. The last element of the scanned set $Lacc$ represents the number of points retained in the current iteration. Its value serves as a criterion to determine whether the iteration will execute again or terminate. If the value is larger than $0$, it means that new feature points should be retained leading to new curve segments generated. Otherwise there is no newly-found point in this iteration and this step terminates.
	%
	% it is difficult to understand "it means that new points could be found, that is, the trajectory is not simplest."
	%
	\item[e.] \textit{Synthesize all of the retained points}. The existing points are firstly copied from $T^{'}$ to $T^{*}$. We then orderly insert the new retained points to $T^{*}$ based on $Is$ and $I^{\max}$. The main steps have been clearly illustrated in Algorithm \ref{alg:updateSet}.
	\item[f.] \textit{Update $Lp$}. As shown in Fig. \ref{Fig05_DPParallel}(d), curve segments are newly generated in above steps, the set $Lc$ should also be changed accordingly. Meanwhile, $Lp$ of each timestamped point can be updated according to the sets $Is$ and $Lacc$ simultaneously. For example, if a feature point is generated related to the $i$-th curve segment, $I_{i}^{'}$ will record the index of start point of the new curve segment. Conditions of the timestamped points along both sides of the feature point can then be updated through the following equation
	\begin{equation}\label{expression11}
	L_{n} =
	\begin{cases}
	L_{n} + L_{i}^{*}& \text{if $n \leq I_{i}^{'}$}, \\
	L_{n} + L_{i}^{*} + 1& \text{otherwise},
	\end{cases}  \\
	\end{equation}
	where $n \in \{1,2, \cdots, N_{o}\}$ and $i \in \{1, 2, \cdots, I_{o}\}$.
	\item[g.] \textit{Return to Step a}. All auxiliary sets are reseted to save storage space, which can facilitate the next generation. In addition, all elements are copied from $T^{*}$ to $T^{'}$ for next iteration.
\end{enumerate}

The flowchart of our proposed GPU-based parallel compression framework is shown in Fig. \ref{Fig05_DPParallel}. It is obviously different from the original DP-based compression method shown in Fig. \ref{Fig04_DPProcessing}. The effectiveness of our parallel compression framework will be demonstrated in Section \ref{sec:TraCompress}.
\begin{algorithm}[t]
	\caption{Matrix-Based Segmented Scan} \label{alg:segmentScan}
	\begin{algorithmic}[1]
		\Require $Lp$, $D$
		\Ensure $ D^{\max}$
		\Comment {/* Step 1. Scan rows using $H$ threads. */}
		\If {threadId $< H$}
		\Comment  {//Copy elements from $D$ to shared memory $Ds$}
		\State T *row = $\&Ds$[threadId * W];
		\State curmaxdist = row[0];
		\State curlabel = $Lp$[0];
		\For {i = 1:$W$}
		\If {$Lp$[i] $!=$ curlabel $||$ row[i] $>$ curmaxdist}
		\State curmaxdist = row[i];
		\State curlabel = $Lp$[i];
		\Else
		\State row[i] = curmaxdist;
		\EndIf
		\State column[threadId] = curmaxdist;
		\State columnLabel[threadId] = curlabel;
		\EndFor
		\EndIf
		\State sync();
		\Comment {/* Step 2. Scan row results based on Step 1 using one thread. */}
		\State scanColumn();
		\State sync();
		\Comment {/* Step 3. Fix rows using $H$ threads. */}
		\If {threadId $< H$}
		\State row = $\&Ds$[threadId * W];
		\State curmaxdist = column[threadId - 1];
		\State curlabel = columnLabel[threadId - 1];
		\For {i = 1:$W$}
		\If {$Lp$[i] $==$ curlabel $\&\&$ row[i] $<$ curmaxdist}
		\State row[i] $=$ curmaxdist;
		\Else
		\State break;
		\EndIf
		\EndFor
		\EndIf
		\Comment //Copy maixum VEDs from $Ds$ to  $D^{\max}$
	\end{algorithmic}
\end{algorithm}
\subsubsection{Scan Algorithm}
\label{Sec:ScanSection}
This algorithm is adopted in DP parallelization to obtain $Lacc$ by scanning $Lc$. Th scanned result $Lacc$ determines whether the iteration will execute again or terminate. If the last element of $Lacc$ is $0$, it means that there is no point needed to retain and the compression has been done. The scan algorithm exploits an algorithmic pattern (i.e., balanced trees) that often arises in the GPU-based parallel computing. The idea behind it is to construct a balanced binary tree on the input data (i.e., $Lc$ in this paper) and sweep it to and from the root to compute the prefix sum. This work-efficient scan algorithm mainly consists of two phases, i.e., the reduce phase (a.k.a., up-sweep phase) and the down-sweep phase \cite{CUDA2019}.
\begin{figure}[t]
	\centering
	\includegraphics[width=1.0\linewidth]{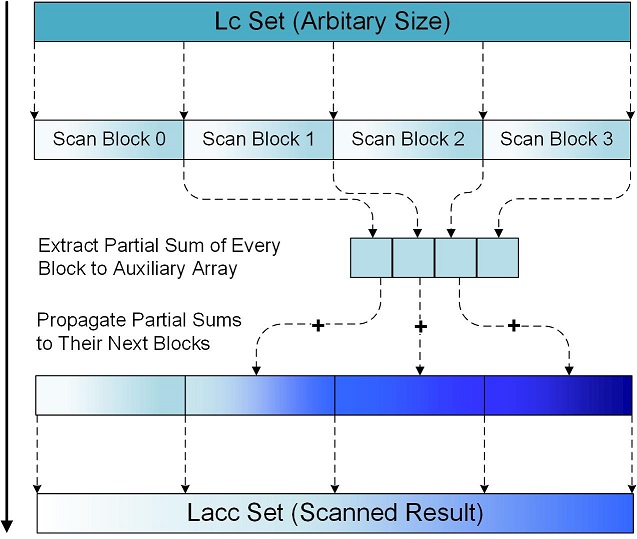}
	\caption{Flowchart of implementing exclusive scan algorithm on one vessel trajectory of arbitrary length.}
	\label{Fig06_Scan}
\end{figure}

In the reduce phase, we traverse the balanced binary tree from leaves to root node (i.e., the last node in $Lc$) while computing partial sums at internal nodes of the tree. Each internal node holds the sum of the partial leaf nodes. After this phase, value of the root node naturally equals the sum of all nodes in $Lc$. In the down-sweep phase, we retrace the tree from the root node to the leaf node. The root node of the tree is first assigned a value of $0$. According to the idea of step-by-step exchange, the zero value of the root node is then exchanged to the first element of $Lacc$. The partial sums from the reduce phase will be summed again and exchanged to build the scan in place on $Lacc$.

Note that the above parallel algorithm is designed for only single thread block. To meet the requirement of simultaneously handling large-scale trajectories, it is necessary to deploy data to multiple blocks. However, data can not be shared between thread blocks. To maintain the integrity of the final results, the last elements of each block, except for the last thread block, are stored in a temporary set. We then launch another kernel function in which elements of the temporary set are propagated to their next blocks. The values of elements in each block are updated again, except for the first block. The flowchart of our method is shown in Fig. \ref{Fig06_Scan}. The final result $Lacc$ will be accordingly updated after all blocks are synthesized.
%
%It should be noted that the values of the last elements of $Lc$ in each block are easy to neglect because of the exclusive scan algorithm introduced. For instance, there are two blocks performing scan algorithm, i.e., $[0, 1, 0, 1]$ and $[0, 1, 1, 0]$. The scanned results of each block are then given, respectively, by $[0, 0, 1, 1]$ and $[0, 0, 1, 2]$. Result of the second block is synthesized as $[1, 1, 2, 3]$. It is obvious that this result ignores the value of the last element in the first block resulting in inaccurate computation. Therefore, before we save the last element of each thread block, it is necessary to check its value to avoid the erroneous final results.
%
\begin{figure*}[htbp]
	\centering
	\includegraphics[width=1.0\linewidth]{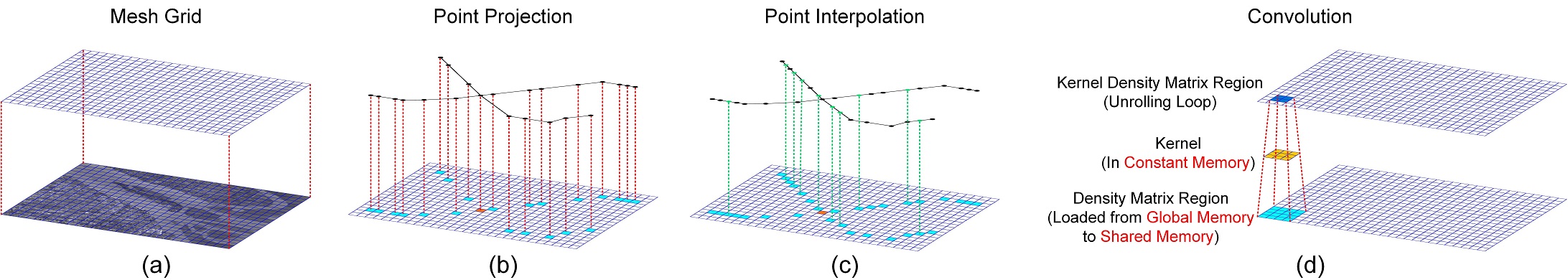}
	\caption{Flowchart of the proposed GPU-accelerated trajectory visualization. We first divide the selected water area into a finite number of grids in (a) to generate a density matrix, then perform parallel data projection in (b). The parallel interpolation in (c) will be implemented if the projected trajectory is discontinuous. The parallel convolution is finally performed on the density matrix to generate trajectory visualization results in (d).}
	\label{Fig07_KDEParallel}
\end{figure*}
\subsubsection{Segmented Scan Algorithm}
\label{sec:SegScan}
Several effective segmented scan algorithms have been currently introduced for GPUs \cite{Scan2007,GPGPU2005,prefixsum2006}. To avoid memory bank conflicts as much as possible \cite{CUDA2019}, we leverage the idea of matrix scan proposed in \cite{Fastscan2008} to concurrently scan the VED of all timestamped points on each curve segment, and then obtain the points with maximum VED of each segment. To support the efficient segmented scan algorithm for large sequences, the set $D$ is deployed to multiple blocks. The set $D$ with length $N_o$ can be divided into ${N_o}/{B}$ sub-sequences, where $B$ is the size of the thread block. Each sub-sequence is then rearranged by a two-dimensional matrix of size $B = W \times H$. In order to fully utilize parallel performance of GPU, the size of $H$ should be a multiple of the warp size. In this paper, the $1080$Ti GPU utilized in our experiments has a warp size of $32$. We thus scan $D$ in a single block by performing the following three steps:

\begin{enumerate}
\item[a.] $H$ threads work in parallel. Each thread scans a row of elements to obtain the maximal VED. Note that $Lp$ is used to distinguish different curve segments. The maximal VED of each row will be stored in auxiliary array \textsf{column} with length of $H$. Meanwhile, their corresponding label values of $Lp$ will be stored in auxiliary array \textsf{columnLabel}.

\item[b.] To handle the case where the timestamped points in the same curve segment are separated by rows, one thread is adopted to scan the \textsf{column} with length $H$. The maximum value of each row will be updated accordingly.

\item[c.] $H$ threads once again scan the matrix in parallel, and update the maximal VED of each segment according to \textsf{column} and \textsf{columnLabel}.
\end{enumerate}

Our implementation of segmented scan algorithm on a single block is detailedly described in Algorithm \ref{alg:segmentScan}. To deal with the case where the points in the same curve segment are separated by blocks, we adopt the scan-recursion-propagate (sRp) approach which is implemented in \cite{Fastscan2008} to handle large-scale sequences. The intermediate results of each block are collected by auxiliary set. We then recursively scan these intermediate results until the scanned results can be processed by one block. The scanned results in multiple blocks will be synthesized in the propagation phase with the final result set $D^{\max}$ being updated accordingly.
\subsection{Parallelization of KDE}
\label{sec:KDEparallelization}
In this subsection, we discuss the specific parallelization of KDE algorithm for visualization of large-scale vessel trajectories. To maximize the degree of parallelism, all trajectories are concurrently processed in our KDE parallelization framework, which is significantly different from our DP parallelization version proposed in Section \ref{sec:PallaDP}. Two main problems should be considered during KDE-based trajectory visualization, including (1) boundaries between different trajectories are obviously complicated; (2) there exists a conflict when numerous threads concurrently execute projection since all timestamped points are projected to the same density matrix $\mathcal{M}_{D}$. In this work, the label mechanism and atomic operation will be exploited to handle the above problems. Our GPU-based parallel visualization framework thus has the capacity of simultaneously visualizing the large-scale vessel trajectories. It is tremendously different from the traditional KDE which visualizes the spatio-temporal trajectory one by one. The flowchart of our parallel visualization framework is illustrated in Fig. \ref{Fig07_KDEParallel}.
\subsubsection{Parallel Projection and Interpolation}
In Section \ref{sec:KDETraj}, we have discussed how to project the timestamped points into the same density matrix $\mathcal{M}_{D}$. For GPU-based parallel projection, it is necessary to consider the write-after-write (WAW) data hazard problem. This WAW hazard commonly occurs when two threads attempt to write data to the same memory location of $\mathcal{M}_{D}$. To effectively handle it, we leverage the atomic operation \cite{CUDA2019} to serialize contentious updates from multiple threads. In particular, the atomic operation permits only one thread to access the same memory address at the same time, which can assist in yielding the correct projection.

We now give the details about the linear interpolation in parallel projection. Let $X_\mathrm{set}$ and $Y_\mathrm{set}$, respectively, store the abscissa and ordinate information of all timestamped points in vessel trajectories. How to unambiguously classify each trajectory is of great importance when we perform data interpolation for generating $\mathcal{M}_{D}$. We thus introduce a label set $Lt$ representing the vessel trajectories. For the same trajectory, all timestamped points share the same label value. In contrast, the points which belong to different trajectories have the different label values. During our visualization experiments, if several timestamped points belong to the same trajectory, we will check whether data interpolation is implemented after these points are projected into $\mathcal{M}_{D}$.

There are two adjacent positions $( {x}_{k}^{n}, {y}_{k}^{n} )$ and $( {x}_{k}^{n+1}, {y}_{k}^{n+1} )$, which denote the Cartesian coordinates of the $n$- and $(n+1)$-th timestamped points in the $k$-th vessel trajectory, respectively. According to Eqs. (\ref{expression5}) and (\ref{expression6}), their projected versions can be easily obtained in $\mathcal{M}_{D}$, i.e., $( \tilde{x}_{k}^{n}, \tilde{y}_{k}^{n} )$ and $( \tilde{x}_{k}^{n+1}, \tilde{y}_{k}^{n+1} )$. Let $\left| \tilde{x}_{k}^{n+1} - \tilde{x}_{k}^{n} \right| = n_x$ and $\left| \tilde{y}_{k}^{n+1} - \tilde{y}_{k}^{n} \right| = n_y$, if $\max\left( n_x, n_y \right) > 1$, the original trajectory $T$ can be assumed to be discontinuous in $\mathcal{M}_{D}$ in this work. To improve the visualization performance, we directly adopt the simple linear interpolation method to reconstruct the missing elements between $( \tilde{x}_{k}^{n}, \tilde{y}_{k}^{n} )$ and $( \tilde{x}_{k}^{n+1}, \tilde{y}_{k}^{n+1} )$ in $\mathcal{M}_{D}$. Let $c_{\max} = \max\left( n_x, n_y \right)$, the missing element $( \tilde{x}_{k}^{n,c}, \tilde{y}_{k}^{n,c})$ with $c=1, 2, \cdots, c_{\max}-1$ can then be calculated as follows
%
% According to Eqs. (\ref{expression5}) and (\ref{expression6}), the position $( \tilde{x}_{k}^{n}, \tilde{y}_{k}^{n} )$ is obtained by projecting the Cartesian coordinate of the $n$-th timestamped point in the $k$-th vessel trajectory into the density matrix $\mathcal{M}_{D}$.
%
\begin{equation}\label{Eq:InterX}
    \tilde{x}_{k}^{n,c} = \left[ \tilde{x}_{k}^{n} + \frac{c}{c_{\max}} \times \left( \tilde{x}_{k}^{n+1} - \tilde{x}_{k}^{n} \right) \right],
\end{equation}
\begin{equation}\label{Eq:InterY}
    \tilde{y}_{k}^{n,c} = \left[ \tilde{y}_{k}^{n} + \frac{c}{c_{\max}} \times \left( \tilde{y}_{k}^{n+1} - \tilde{y}_{k}^{n} \right) \right],
\end{equation}
with $\tilde{x}_{k}^{n} \leq \tilde{x}_{k}^{n,c} \leq \tilde{x}_{k}^{n+1}$ and $\tilde{y}_{k}^{n} \leq \tilde{y}_{k}^{n,c} \leq \tilde{y}_{k}^{n+1}$. Here, $\left[ \cdot \right]$ denotes the rounding operation. Since all threads concurrently perform interpolation in our GPU-based parallel framework, the efficiency of parallelization can be greatly improved. It can thus solve the problem of low efficiency existed in CPU-based serial interpolation. Our implementation of parallel projection on a single block is described in Algorithm \ref{alg:ProjectionKernel}. It is worth noting that interpolation should be divided into two conditions due to the uncertain directions of vessel trajectories.

\subsubsection{Parallel Convolution}
It is well known the elements in the density matrix $\mathcal{M}_{D}$ are independent of each other. The convolution operation, related to KDE visualization, can be independently and parallelly executed for all elements $(\tilde{x}, \tilde{y})$ with $1 \leq \tilde{x} \leq u$ and $1 \leq \tilde{y} \leq v$ in $\mathcal{M}_{D}$. To effectively parallelize the convolution operations using GPU, $\mathcal{M}_{D}$ will be divided into blocks by a thread grid. Inspired by \cite{2Dconvolution2013}, our parallel convolution algorithm first leverages the adjacent threads to cooperatively pre-load several blocks of $\mathcal{M}_{D}$ into the shared memory. The convolution kernel is then stored in the constant memory. In our experiments, we compute a region of output pixels using unrolling loops. The unrolling loops have become a common technique which can reduce the computational cost of branch prediction. Each thread can produce more pixels with unrolling loops at one iteration, which can further promote the instruction level parallelism (ILP). We finally write the output pixels back to the global memory. With the size of convolution kernel becomes larger, our strategy can produce higher efficiency and more robust performance compared with conventional parallel libraries.
\begin{figure}[t]
	\centering
	\includegraphics[width=1.0\linewidth]{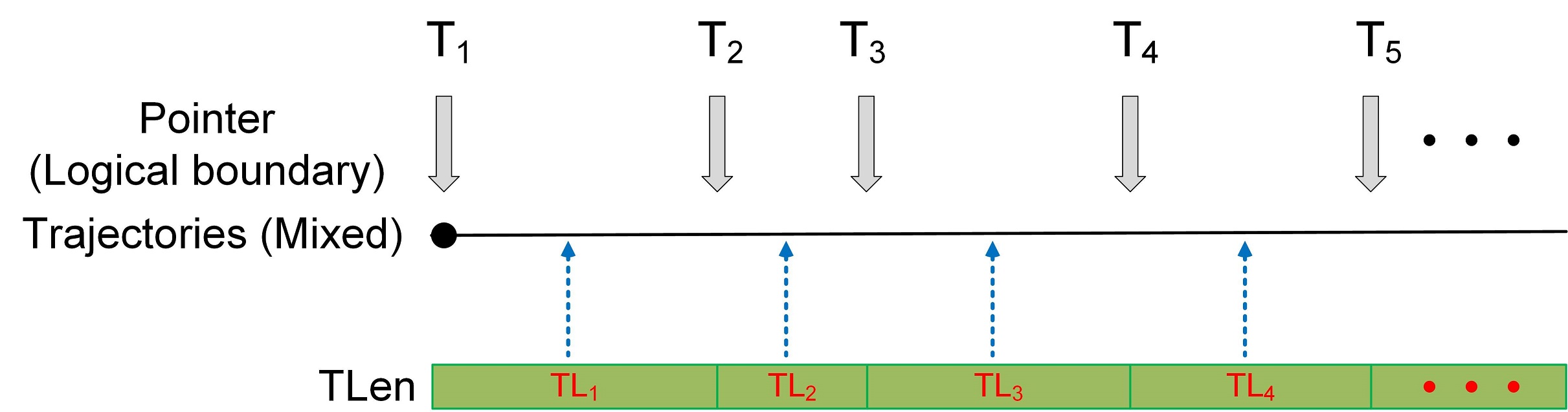}
	\caption{Coalesced global memory access.}
	\label{Fig08_MemoryAccess}
\end{figure}
\begin{algorithm}[t]
	\caption{Generation of Density Matrix} \label{alg:ProjectionKernel}
	\begin{algorithmic}[1]
		\Require $X_\mathrm{set}$, $Y_\mathrm{set}$, $Lt$
		\Ensure density matrix ($\mathcal{M}_{D}$)
		\State int tid = blockId * blockDim + threadId;
		\Comment // Allocate Shared Memory to accelerate Computing
		\State \_\_shared\_\_ T sX[blockDim];
		\State T *sY = sX + blockDim;
		\Comment /* Step 1. Compute coordinate in $\mathcal{M}_{D}$ for each point based on Eq.\ref{expression5} and Eq.\ref{expression6} */
		\State sX[threadId] = CoordinateConversion($X_\mathrm{set}$[tid]);
		\State sY[threadId] = CoordinateConversion($Y_\mathrm{set}$[tid]);
		\State sync();
		\Comment /* Step 2. Project converted coordinates to $\mathcal{M}_{D}$. */
		\State atomicAdd(\&$\mathcal{M}_{D}$[sY[threadId] $\times$ $u$ + sX[threadId]], 1);
		\Comment /* Step 3. Linear interpolation based on Eq.\ref{Eq:InterX} and Eq.\ref{Eq:InterY} */
		\If {$Lt$[tid] == $Lt$[tid + 1]}
		\State nx = abs(sX[threadId +1] - sX[threadId]);
		\State ny = abs(sY[threadId +1] - sY[threadId]);
		\State $c_{max}$ = max(nx,ny);
		\If { $c_{max}$ $>$ 1}
		\For {c = 1: $c_{max}$}
		\State x = InterpolateX(sX[threadId], $c_{max}$,
		\State     sX[threadId + 1], c);
		\State y = InterpolateY(sY[threadId], $c_{max}$,
		\State     sY[threadId + 1], c);
		\State atomicAdd(\&$\mathcal{M}_{D}$[y $\times$ $u$ + x], 1);
		\EndFor
		\EndIf
		\EndIf
	\end{algorithmic}
\end{algorithm}
\subsection{Memory Access Optimization}
\subsubsection{Compression of Massive Trajectories}
\label{massive data}
Traditional DP algorithm can only compress one vessel trajectory at one time. When dealing with large-scale vessel trajectories, an intuitive method is to allocate global memory for each trajectory at the start and release the memory at the end of compression. It is well known that frequent memory allocation and release are time-consuming, which easily bring negative effects on compression acceleration. To achieve coalesced access to the global memory, all vessel trajectories will be merged in our GPU-based DP parallelization framework. To logically classify the different trajectories, we define an array \textsf{TLen} which is utilized to store the lengths of all input trajectories. As shown in Fig. \ref{Fig08_MemoryAccess}, the whole trajectory dataset, originally fetched from database, will be directly copied from host memory to device memory at one time. Assume that a trajectory is completely compressed, the pointer will move to the head point of its next trajectory. In addition, the memory spaces of all auxiliary sets (e.g., $Lp$, $Lc$, etc.) are allocated only once in our parallelization framework. The memory spaces will be resetted at the beginning of DP algorithm for next compression. Meanwhile, the size of memory spaces is essentially determined by the maximal value of \textsf{TLen}. It not only prevents the memory overflow, but also avoids the frequent allocation and release of video memory. To significantly shorten computational time, our proposed method only performs one memory allocation at the beginning of DP compression, and performs one memory release at the end.
%
% A pointer move on the sets according to \textsf{TLen}.
%
%
\subsubsection{Utilization of On-Chip Memory}
There are two types of memory (i.e., on-board and on-chip memories) that are equipped on the GPU chip \cite{IoTGPU2019}. In particular, the global memory belongs to the on-board memory. The on-chip memory, associated with each SM, includes the shared memory and registers. The major differences between on-board and on-chip memories are their latency and capacity. The minimal latency is the main advantage of the on-chip memory in practice. To minimize the latency in accessing data, especially for the data needed to be accessed repeatedly, this paper adopts the scheme that prioritizes shared memory. Meanwhile, the utilization of registers can be increased while guaranteeing the registers do not overflow and are adopted for storing intermediate variables. Given an example in our parallelization framework, when computing VED for each timestamped point, we first extract the known feature points set from the global memory to the shared memory, and then transfer their first and last points to the registers. This implementation is capable of achieving the goal of minimal latency in memory access.
\begin{figure}[t]
	\centering
	\includegraphics[width=1.0\linewidth]{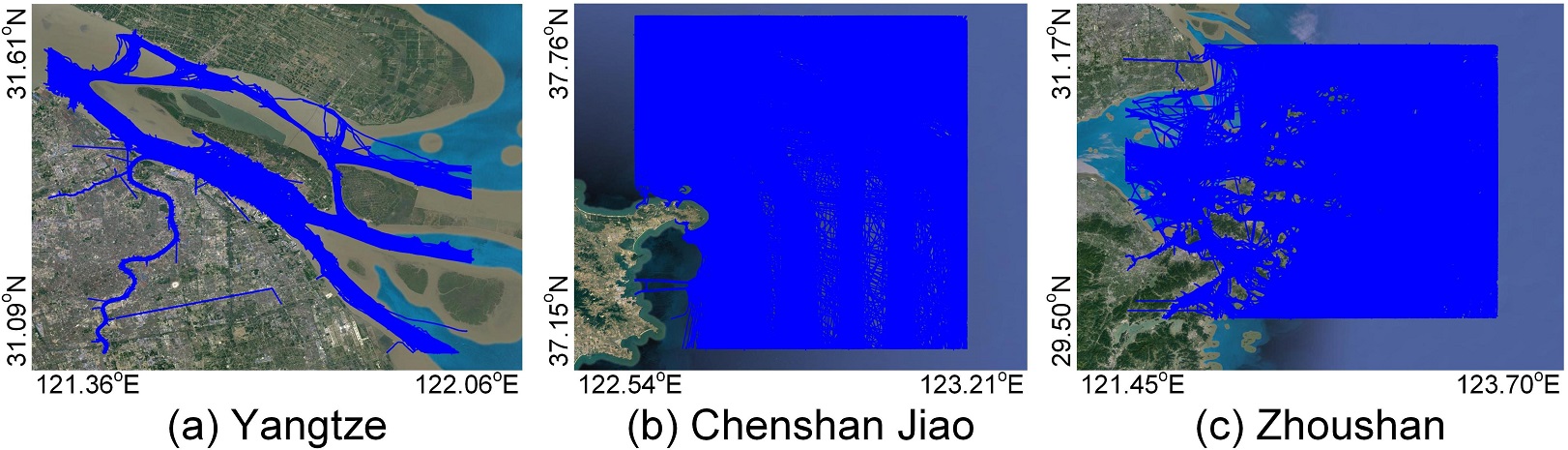}
	\caption{Realistic AIS-based vessel trajectories, utilized in our experiments, are collected from: (a) South Channel of Yangtze River Estuary, (b) Chengshan Jiao Promontory, and (c) Zhoushan Islands, respectively.}
	\label{Fig09_OriTraj}
\end{figure}
\section{Experimental Results and Discussion}
\label{section4}
To evaluate the effectiveness and efficiency of the proposed GPU-based parallelization frameworks, they will be compared with CPU-based serial implementations on realistic large-scale vessel trajectories. In particular, experimental results on DP-based trajectory compression are parallelly implemented to select the optimal compression threshold. Parallel visualization experiments are also generated to determine the most suitable KDE algorithm for vessel density mapping. The source code is available at \url{http://mipc.whut.edu.cn}.
%
%Meanwhile, to guarantee a good performance for both vessel trajectory compression and visualization, experiments are conducted to select the appropriate threshold for DP-based compression and kernel function as well as kernel size for KDE-based data visualization, which can achieve satisfactory visualization of vessel trajectory data visualization.
%
\subsection{Experimental Settings}
\subsubsection{Experimental Environment}
To illustrate the satisfactory performance of our parallelization frameworks, the essential experimental environments should be established for the parallel implementations. Both software and hardware environments are detailedly illustrated in Table \ref{Table2}. The compression and visualization of large-scale vessel trajectories in this work will be implemented accordingly.
\begin{table}[t]
	\centering
	\caption{The hardware and software environments.}
	\begin{tabular}{|c|c||c|c|}
		\hline
		Hardware      & Model   & Software  & Version \\ \hline \hline
		CPU           & \tabincell{c}{I$7$-$7700$K\\Quad Core}&  \tabincell{c}{Microsoft\\Visual Studio} & $2017$ \\ \hline
		Host Memory   & $16$GB DDR$4$     & CUDA          & $10.0$      \\ \hline  %\cline{1-2} %\cline{4-5}
		GPU           & GTX $1080$Ti      & Mysql         & $8.0$       \\ \hline %\cline{1-2} %\cline{4-5}
		Global Memory & $11$GB GDDR$5$    & $\setminus$   & $\setminus$ \\ \hline
	\end{tabular} \label{Table2}
\end{table}
\begin{table*}[t]
	\centering
	\caption{The statistical and geometrical information related to $3$ different water areas. The boundary points are in the form of geographical coordinates.\label{RegionInfor}}
	\begin{tabular}{|c||c|c|c|c|c|}
		\hline
		Water Areas                                     & \tabincell{c}{Number of\\Vessel Trajectories} & \tabincell{c}{Number of\\Timestamped Points}          &\tabincell{c}{Boundary\\Points} & Longitude$(^\circ)$ & Latitude$(^\circ)$ \\ \hline \hline
		\multirow{2}{*}{\tabincell{c}{South Channel of\\Yangtze River Estuary}} & \multirow{2}{*}{$48272$}        & \multirow{2}{*}{$55069187$} & Left Top       & $121.3795$  & $31.5746$  \\ %\cline{4-6}
		&                               &                           & Right Bottom   & $121.9842$  & $31.1166$  \\ \hline
		\multirow{2}{*}{\tabincell{c}{Chengshan Jiao\\Promontory}}              & \multirow{2}{*}{$28010$}        & \multirow{2}{*}{$18979621$} & Left Top       & $122.5833$  & $37.7500$  \\ %\cline{4-6}
		&                               &                           & Right Bottom   & $123.1667$  & 37.1667  \\ \hline
		\multirow{2}{*}{Zhoushan Islands}                       & \multirow{2}{*}{$18623$}        & \multirow{2}{*}{$15521563$} & Left Top       & $121.5056$  & $31.0993$  \\ %\cline{4-6}
		&                               &                           & Right Bottom   & $123.6127$  & $29.5607$  \\ \hline
	\end{tabular}
\end{table*}
\subsubsection{Experimental Datasets}
The original vessel trajectories were collected from the terrestrial AIS base stations in $3$ different water areas (i.e., the South Channel of Yangtze River Estuary, the Chengshan Jiao Promontory, and the Zhoushan Islands), visually illustrated in Fig. \ref{Fig09_OriTraj}. The statistical and geometrical information related to these selected water areas are detailedly shown in Table \ref{RegionInfor}. In our experiments, the Maritime Mobile Service Identify (MMSI) and time stamps were utilized to classify the collected AIS-based vessel trajectories. It is well known that the collected raw AIS data often suffer from undesirable outliers during signal acquisition. To eliminate the negative influences of these outliers on trajectory computations, the qualities of original vessel trajectories have been improved using the effective methods introduced in our previous studies \cite{LiLiuWCSP2016,LiangLiuICBDA2019}.
\subsection{Performance Indexes on Trajectory Compression}
To quantitatively evaluate the trajectory compression performance, four different quality measures were simultaneously utilized, i.e., Compression Ratio (CR), Rate of Length Loss (RLL), Dynamic Time Warping (DTW), and Speedup Ratio (SR). In particular, CR reflects the compression degree. Both RLL and DTW are utilized to comprehensively evaluate the information integrity of compressed trajectories. Since our contributions mainly focus on the parallelizations of DP and KDE algorithms, SR is also considered to evaluate the computational cost. The mathematical definitions are given as follows
%
%These performance indexes on DP-Based data compression adopted in our experiments are defined as follow:
%
\subsubsection{Compression Ratio (CR)}
CR has become the most common compression index which can describe the volume change of trajectory data. It can be defined as follows
\begin{equation}\label{expressionCR}
    \mathrm{CR} = \left( 1 - \frac{N_c}{N_o} \right) \times 100 \%,
\end{equation}
where ${N_c}$ and ${N_o}$, respectively, denote the numbers of timestamped points in compressed and original vessel trajectories.
%
%\begin{figure}[t]
%	\centering
%	\includegraphics[width=0.8\linewidth]{Fig10_DTW.jpg}
%	\caption{The visual illustration of DTW algorithm. From left to right: (a) displays the DTW-based alignments between original vessel trajectory $T$ and its compressed version $T'$ for measuring distance (or similarity), and (b) can show the warping paths, generated by calculating DTW algorithm. It can be easily found that DTW algorithm can calculate the similarity of two vessel trajectories with different lengths.}
%	\label{FigDTW}
%\end{figure}
%
\subsubsection{Rate of Length Loss (RLL)}
RLL essentially reflects the rate of the length loss and the total length of original trajectories. Similar with the definitions in Section \ref{sec:methods}, we define one original vessel trajectory $T$ as $T = \left\{ P_{1}, P_{2}, \cdots, P_{N_o} \right\}$ with $N_o$ being the number of timestamped points. The compressed vessel trajectory $T'$ is defined as $T' = \left\{ P'_{1}, P'_{2}, \cdots, P'_{N_c} \right\}$ with $N_c \ll N_o$. Furthermore, the original vessel trajectories and their compressed versions can be represented, respectively, by $\left\{ T_{1}, T_{2}, \cdots, T_{M} \right\}$ and $\left\{ T'_{1}, T'_{2}, \cdots, T'_{M} \right\}$ with $M$ denoting the total number of vessel trajectories utilized in our experiments. The definition of RLL is given by
%
%\begin{equation}\label{expression12}
%    \left| T \right| = \sum _{n=1}^{N_o-1} \left| P_{n} P_{n+1} \right|,
%\end{equation}
%
%\begin{equation}\label{expression13}
%    \left| T' \right| = \sum _{n=1}^{N_c-1} \left| P'_{n} P'_{n+1} \right|,
%\end{equation}
%
%\begin{equation}\label{expression14}
%    \mathrm{LL} = \sum _{m=1}^{M} \left| T_{m} \right| - \sum _{m=1}^{M} \left| T'_{m} \right|,
%\end{equation}
%
\begin{equation}\label{expression15}
    \mathrm{RLL} = \mathrm{LL} / \sum _{m=1}^{M}|T_{m}|,
\end{equation}
where $\mathrm{LL} = \sum _{m=1}^{M} \left| T_{m} \right| - \sum _{m=1}^{M} \left| T'_{m} \right|$ with $\left| T \right| = \sum _{n=1}^{N_o-1} \left| P_{n} P_{n+1} \right|$ and $\left| T' \right| = \sum _{n=1}^{N_c-1} \left| P'_{n} P'_{n+1} \right|$. Here, $\left| T \right|$ and $\left| T' \right|$, respectively, denote the lengths of one original trajectory and its compressed version. Trajectory length is defined as the sum of the distance between two adjacent timestamped points in one trajectory. $\mathrm{LL}$ denotes the difference between the total lengths of original and compressed trajectories.
\subsubsection{Dynamic Time Warping (DTW)}
\begin{table*}[!]
	\centering
	\caption{DP-accelerated compression results for different vessel trajectory datasets collected from $3$ different water areas (from left to right: South Channel of Yangtze River Estuary, Chengshan Jiao Promontory, and Zhoushan Islands, respectively).\label{Table4}}
	\begin{tabular}{|c||c|c|c|c|}
		\hline
		Threshold (m) & Number of Timestamped Points & CR (\%)  & RLL (\%)   & DTW ($\mu \pm \delta$)  \\ \hline\hline
		$0.0$          & $55069187/18979621/15521563$      & $0.000 / 0.000 / 0.000   $ & $0.000 / 0.000 / 0.000$ & $0.00 \pm 0.00 / 0.00 \pm 0.00 / 0.00 \pm 0.00$ \\ \hline
		$0.1$          & $22442370/15929948/13830378$      & $59.25 / 16.07 / 10.90$ & $0.018 / 0.001 / 0.004$ & $0.11 \pm 0.13 / 0.07 \pm 0.16 / 0.04 \pm 0.09$ \\ \hline
		$0.5$          & $12922299/10485858/10290482$      & $76.53 / 44.75 / 33.70$ & $0.099 / 0.002 / 0.015$ & $0.31 \pm 0.31 / 0.23 \pm 0.23 / 0.19 \pm 0.27$ \\ \hline
		$1.0$          & $\ \,8904204/\ \,7292010/\ \,7954963$         & $83.83 / 61.58 / 48.75$ & $0.182 / 0.005 / 0.027$ & $0.52 \pm 0.49 / 0.44 \pm 0.37 / 0.36 \pm 0.51$ \\ \hline
		$5.0$          & $\ \,3097010/\ \,2284691/\ \,3526843$         & $94.38 / 87.96 / 77.28$ & $0.416 / 0.030 / 0.089$ & $1.88 \pm 1.91 / 2.13 \pm 1.73 / 1.67 \pm 2.41$ \\ \hline
		$10.0$         & $\ \,1968923/\ \,1284330/\ \,2316715$         & $96.43 / 93.23 / 85.07$ & $0.507 / 0.050 / 0.139$ & $3.01 \pm 3.06 / 3.98 \pm 3.06 / 2.91 \pm 4.14$ \\ \hline
	\end{tabular}
\end{table*}
DTW is a robust distance measure to calculate the similarity (inversely proportional to the distance) between each two time series. Its basic principle is to adopt the dynamic programming approach to find the minimum distance between the time series. Note that each vessel trajectory with several timestamped points is essentially a time series. To evaluate the compression quality, DTW is introduced to measure the similarity between the original trajectory $T = \left\{ P_{1}, P_{2}, \cdots, P_{N_o} \right\}$ and its compressed version $T' = \left\{ P'_{1}, P'_{2}, \cdots, P'_{N_c} \right\}$. If the important feature points are preserved in the compressed version, the corresponding similarity (or distance) will be high (or short). We create a $N_o \times N_c$ grid $\mathcal{D}$ where each grid element $\mathcal{D}_{ s, t }$ along the warping path denotes the accumulated distance between points $P_{s}$ and $P'_{t}$.
%As illustrated in Fig. \ref{FigDTW}, we create a $N_o \times N_c$ grid $\mathcal{D}$ where each grid element $\mathcal{D}_{ s, t }$ along the warping path in Fig. \ref{FigDTW}(b) denotes the accumulated distance between points $P_{s}$ and $P'_{t}$.

Let $\mathbf{P}$ denote a warping path between $T$ and $T'$, which is essentially a sequence $\mathbf{P} = \left\{ \bar{P}_{1}, \bar{P}_{2}, \cdots, \bar{P}_{L} \right\}$ with $\bar{P}_{l} = \left( o_l, c_l \right) \in \left[ 1 : N_o \right] \times \left[ 1 : N_c \right]$. The set of all potential warping pathes is represented by $\mathbb{P}_{N_o N_c}$. The length of a warping path $\mathbf{P}$ satisfies $\max \left( N_o, N_c \right) \leq L \leq N_o + N_c$. The warping cost $d_{\mathbf{P}} \left( T, T' \right)$ of the warping path $\mathbf{P}$ is thus given by
\begin{equation}\label{Eq:WarpingCost}
    d_{\mathbf{P}} \left( T, T' \right) = \sum_{l = 1}^{L} d \left( P_{o_l}, P'_{c_l} \right),
\end{equation}
where $d \left( \cdot, \cdot \right)$ denotes the squared Euclidean distance. The DTW distance between two vessel trajectories $T$ and $T'$, related to the minimum warping cost, can be defined as follows
\begin{align}
    \mathrm{DTW} \left( T, T' \right) & = \sqrt{d_{\mathbf{P^*}} \left( T, T' \right)} \label{expression15} \\
                                      & = \min \left\{ \sqrt{d_{\mathbf{P}} \left( T, T' \right)} ~|~ \mathbf{P} \in \mathbb{P}_{N_o N_c} \right\} \nonumber,
\end{align}
where $\mathbf{P^*}$ denotes the optimal warping path indicating the minimum warping cost. For the sake of better understanding, the accumulated cost matrix $\mathcal{D}$ is introduced to represent the optimal warping path $\mathbf{P^*}$. In particular, each element in $\mathcal{D}$ can be calculated using the following formula
\begin{equation}
	\mathcal{D}_{ i, j } = d \left( P_{i}, P'_{j} \right) + \min \left\{ \mathcal{D}_{ i-1, j-1 }, \mathcal{D}_{ i-1, j }, \mathcal{D}_{ i, j-1} \right\}.
\end{equation}

The optimal warping path $\mathbf{P^*}$ can be generated by back-tracking the accumulated cost matrix $\mathcal{D}$ from $\left( N_o, N_c \right)$ to $(1, 1)$. The calculated DTW distance $\mathrm{DTW} \left( T, T' \right)$ is equivalent to $\sqrt{ \mathcal{D}_{N_o, N_c}}$. The smaller the DTW distance is, the more similar these two trajectories are. Please refer to \cite{LiuTSP2019} for more details on DTW. Note that a tremendous amount of vessel trajectories will be considered in our experiments. The expected value $\mu$ and standard deviation $\delta$ of all DTW distances are thus calculated to evaluate the trajectory compression performance. In particular, the expected value $\mu$ is defined as follows
% expectation
% Then, DTW distance for all vessel trajectories are averaged to ensure the stability of result. The corresponding mathematical expectation $\mu$ is defined as follows:
%
\begin{equation}\label{expression15}
    {\mu = \frac{1}{M} \sum_{m=1}^{M} \mathrm{DTW} \left( T_{m}, T'_{m} \right),}
\end{equation}
with $\mathrm{DTW} \left( T_{m}, T'_{m} \right)$ denoting the DTW distance between the $m$-th original vessel trajectory and its compressed version. In addition, the standard deviation $\delta$ is also selected to illustrate the stability of the DP-based trajectory compression. Its mathematical formula is given by
\begin{equation}\label{expression15}
    \delta = \sqrt{\frac{1}{M} \sum\nolimits_{m=1}^{M} \left[ \mathrm{DTW} \left( T_{m}, T'_{m} \right)- \mu \right]^{2}}.
\end{equation}
\begin{figure*}[t]
	\centering
	\includegraphics[width=1.00\linewidth]{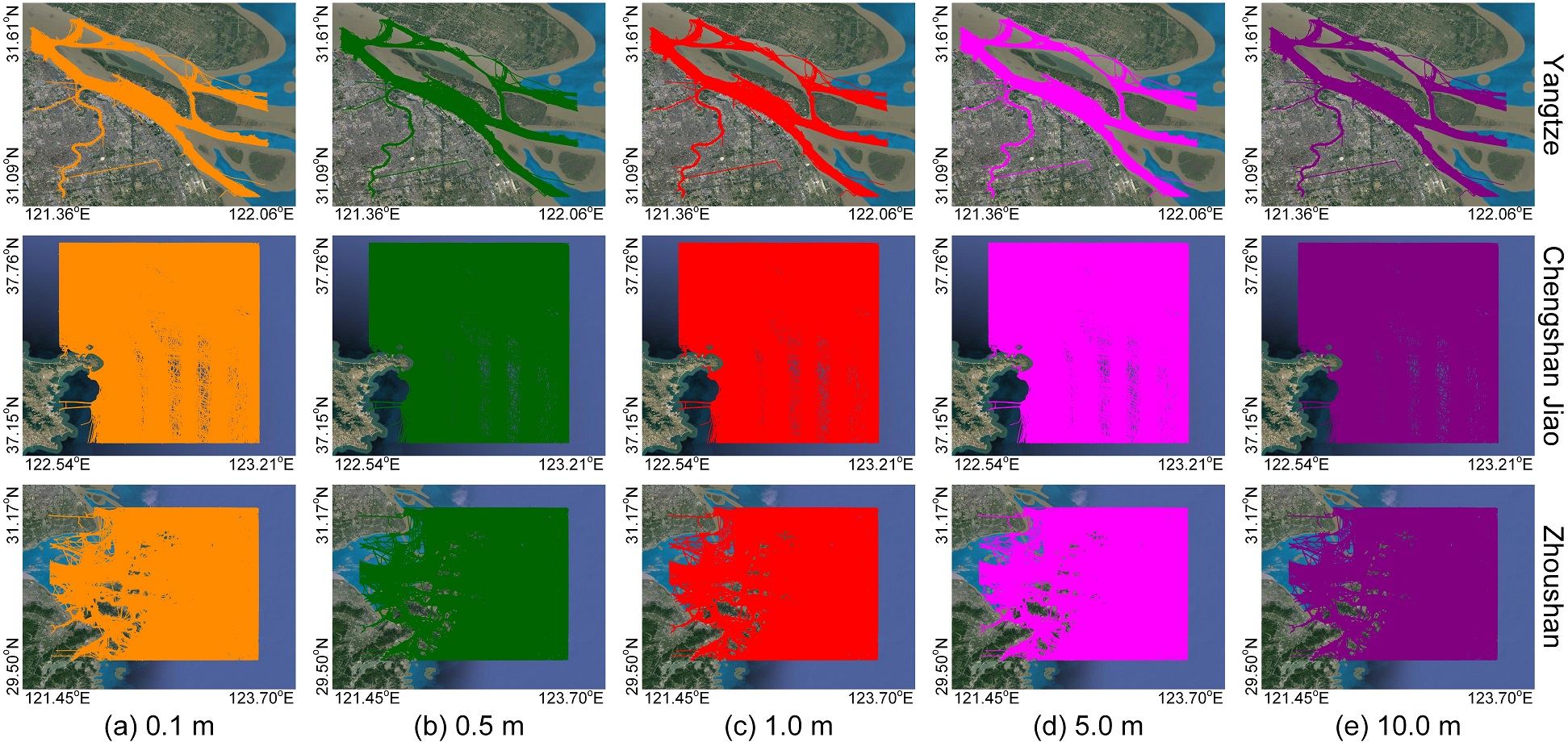}
	\caption{The mappings of compressed vessel trajectories generated with $5$ different DP compression thresholds in $3$ different water areas, i.e., South Channel of Yangtze River Estuary (Top), Chengshan Jiao Promontory (Middle), and Zhoushan Islands (Bottom). From left to right: the DP compression thresholds are (a) $0.1$m, (b) $0.5$m, (c) $1.0$m, (d) $5.0$m, and (e) $10.0$m, respectively.}
	\label{Fig10_CompreTraj}
\end{figure*}
\begin{figure}[t]
	\centering
	\includegraphics[width=1.0\linewidth]{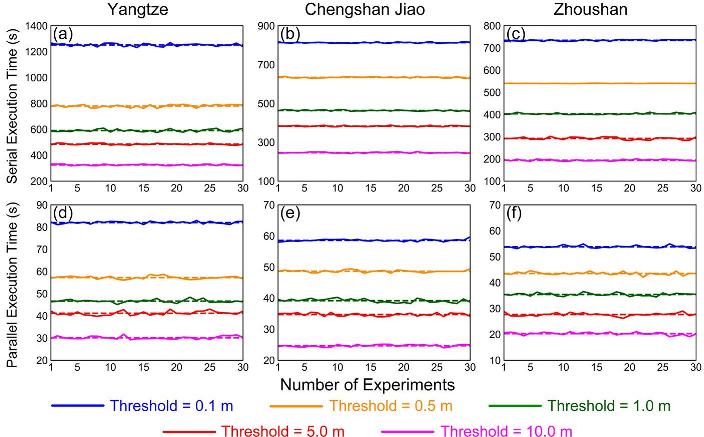}
	\caption{The comparisons of CPU and GPU execution times for $5$ different DP thresholds (i.e., $0.1$m, $0.5$m, $1.0$m, $5.0$m and $10.0$m) in $3$ different water areas. From top to bottom: (a)-(c) display the CPU execution times, and (d)-(f) display the GPU execution times, respectively. The dashed lines in each subfigure indicate the mean execution time.}
	\label{Fig11_DPtime}
\end{figure}
\begin{figure}[!]
	\centering
	\includegraphics[width=1.0\linewidth]{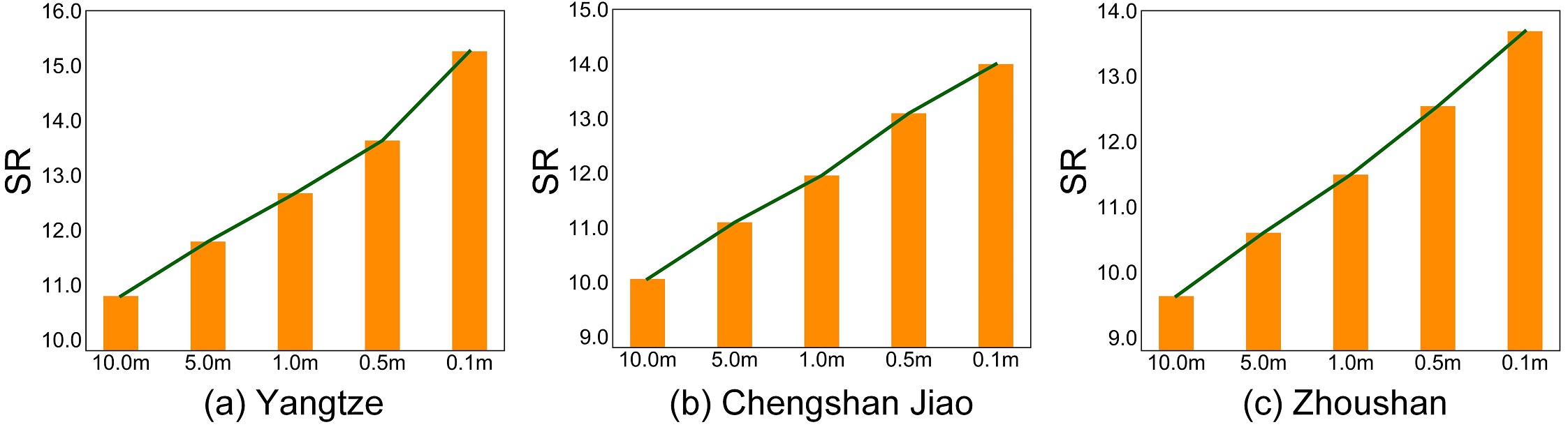}
	\caption{The speedup ratio (SR) of our GPU parallel implementations for $5$ different DP thresholds (i.e., $10.0$m, $5.0$m, $1.0$m, $0.5$m and $0.1$m). From left to right: vessel compression experiments are performed in (a) South Channel of Yangtze River Estuary, (b) Chengshan Jiao Promontory, and (c) Zhoushan Islands, respectively.}
	\label{Fig12_DPratio}
\end{figure}
\begin{figure}[t]
	\centering
	\includegraphics[width=1.0\linewidth]{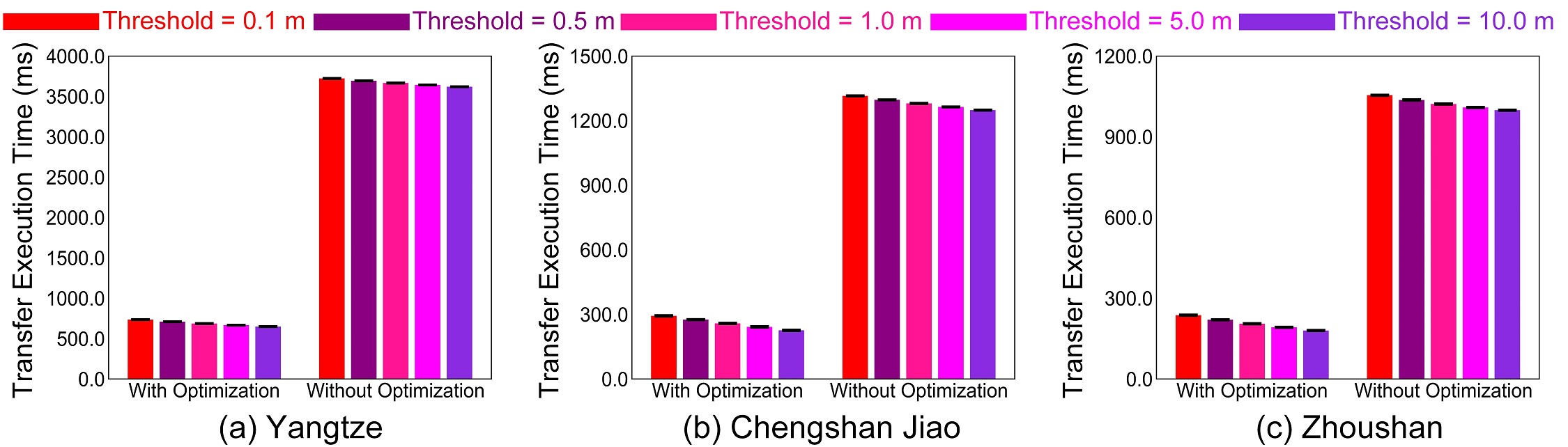}
	\caption{The comparisons of data transfer times with/without memory access optimization. Experiments on data transfer time measurement are implemented for $5$ different DP thresholds in $3$ different water areas.}
	\label{Fig13_OpMemory}
\end{figure}

\begin{figure}[!]
	\centering
	\includegraphics[width=1.0\linewidth]{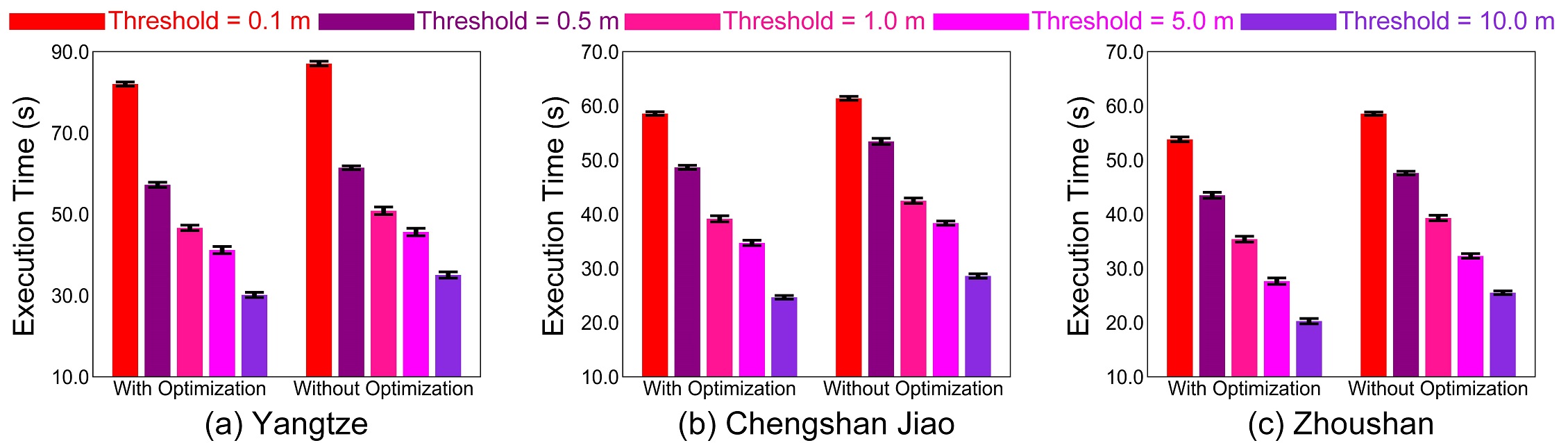}
	\caption{The comparisons of total GPU execution times with/without memory access optimization. Experiments on total GPU execution time measurement are implemented for $5$ different DP thresholds in $3$ different water areas.}
	\label{Fig14_OpTime}
\end{figure}
\subsubsection{Speedup Ratio (SR)}
SR essentially denotes the ratio of the execution time of CPU serial implementation to that of GPU parallel implementation for the same computation task. In our experiments, it is worth noting that the CPU computational cost only includes the execution time of compression or visualization since there is no data transfer in serial implementations. In contrast, the GPU computational cost includes not only the execution time, but also the time of twice data transfers between host and device memories. The data transfer time could lead to significant negative influence on efficient parallel implementations if the GPU codes are not optimal. Especially for the case of large-scale vessel trajectories, this time can not be neglected in the case of calculation of total computational time. It is consequently more practical to consider the data transfer time in GPU parallel implementations in maritime IoT industries.
%
% the ratio of the CPU serial execution time to the GPU parallel execution time.
%
\begin{figure*}[t]
	\centering
	\includegraphics[width=1.00\linewidth]{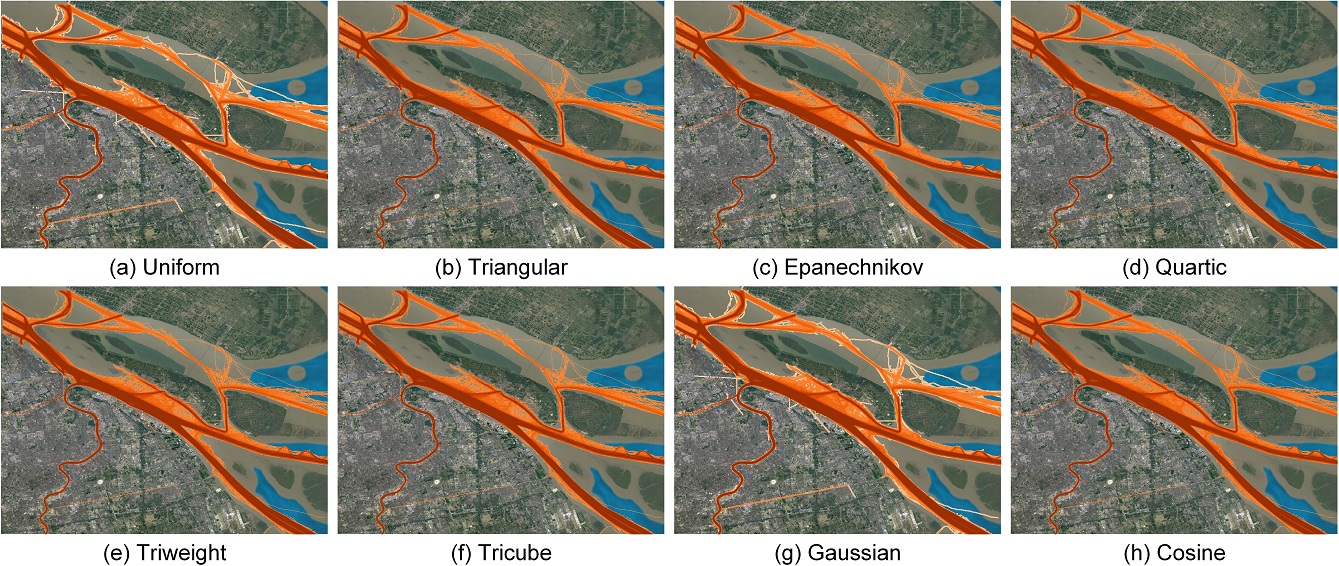}
	\caption{Visualization of large-scale vessel trajectories generated by $8$ different kernel functions of size $3 \times 3$. From top-left to bottom-right: visualization results correspond to (a) Uniform, (b) Triangular, (c) Epanechnikov, (d) Quartic, (e) Triweight, (f) Tricube, (g) Gaussian, and (h) Cosine, respectively.}
	\label{Fig15_Kernel3}
\end{figure*}
\begin{figure*}[t]
	\centering
	\includegraphics[width=1.00\linewidth]{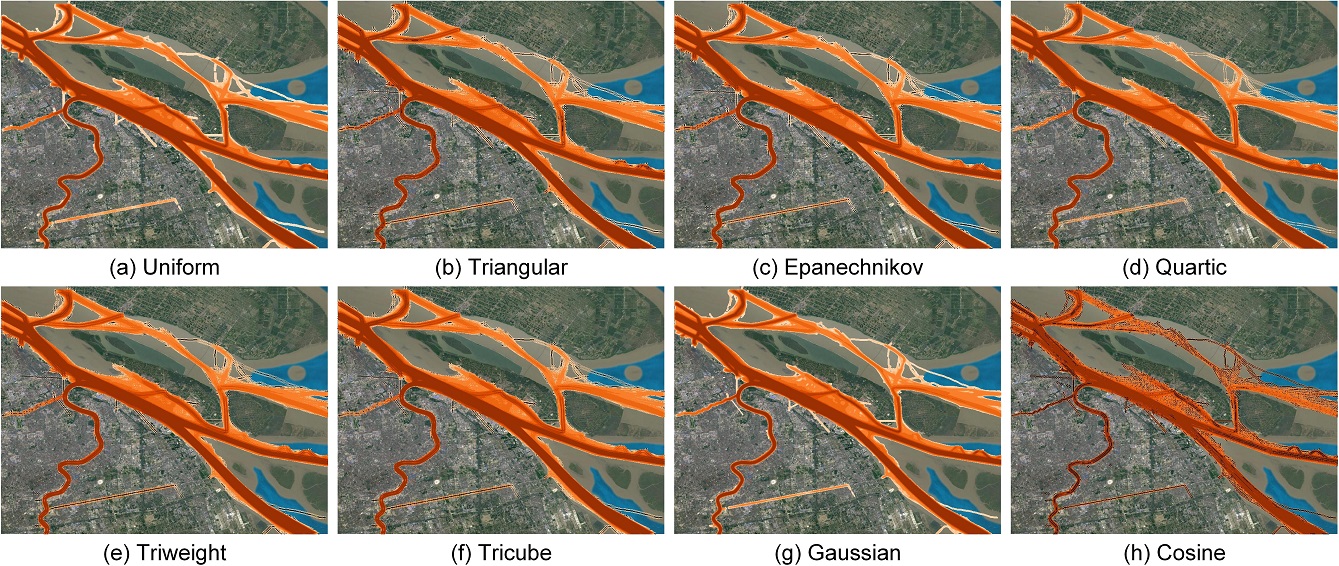}
	\caption{Visualization of large-scale vessel trajectories generated by $8$ different kernel functions of size $7 \times 7$. From top-left to bottom-right: visualization results correspond to (a) Uniform, (b) Triangular, (c) Epanechnikov, (d) Quartic, (e) Triweight, (f) Tricube, (g) Gaussian, and (h) Cosine, respectively.}
	\label{Fig16_Kernel7}
\end{figure*}
\subsection{Experiments on DP-Based Trajectory Compression}
\label{sec:TraCompress}
\subsubsection{Quantitative Evaluation Results}
To quantitatively evaluate the compression performance, CR, RLL and DTW are simultaneously adopted under $6$ different compression thresholds between $0.0$m and $10.0$m. The evaluation results are detailedly illustrated in Table \ref{Table4}. The compression threshold of $0.0$m means there is no difference between original and compressed trajectories. If the high qualities of compressed trajectories are guaranteed, high CR, low RLL and high similarity should be concurrently obtained. As shown in Fig. \ref{Fig10_CompreTraj}, there are only slight differences between the mappings of compressed vessel trajectories for different compression thresholds, however, CR, RLL and DTW distance (i.e., $\mu$) tend to change on different degrees in Table \ref{Table4}. The number of feature points preserved in compressed trajectories is decreased as the compression threshold increases. The CR is accordingly increased to potentially reduce the computational cost during visualization of large-scale vessel trajectories. However, the high RLL and low similarity could result in low-quality visualization in practice. As described in Table \ref{Table4}, if the threshold $\epsilon > 1.0$, both RLL and DTW distance (i.e., $\mu$) will significantly increase in the $3$ different water areas. In contrast, the increase of CR is limited, leading to the degradation of trajectories' quality. It is thus important to select a proper DP threshold to perform trajectory compression. From a practical point of view, significant differences between original and compressed trajectories could cause serious negative influences on trajectory visualization. The influences of different DP thresholds on visualization performance will be detailedly discussed in Section \ref{KDE Experiment}.
%
%Fig.\ref{Figure11} shows the map of simplified trajectory with different compression thresholds.
%
%Regarding to the experimental results on RLL and DTW, both of them also increase with the threshold increases, especially when the compression threshold increases from $1.0$m to $5.0$m, the RLL and $\mu$ increase obviously. We can conduct that the similarity between original data and compressed data decreases with threshold increases. However, the variation of similarity have no effect on trajectory position information.
%
\begin{figure*}[t]
	\centering
	\includegraphics[width=1.00\linewidth]{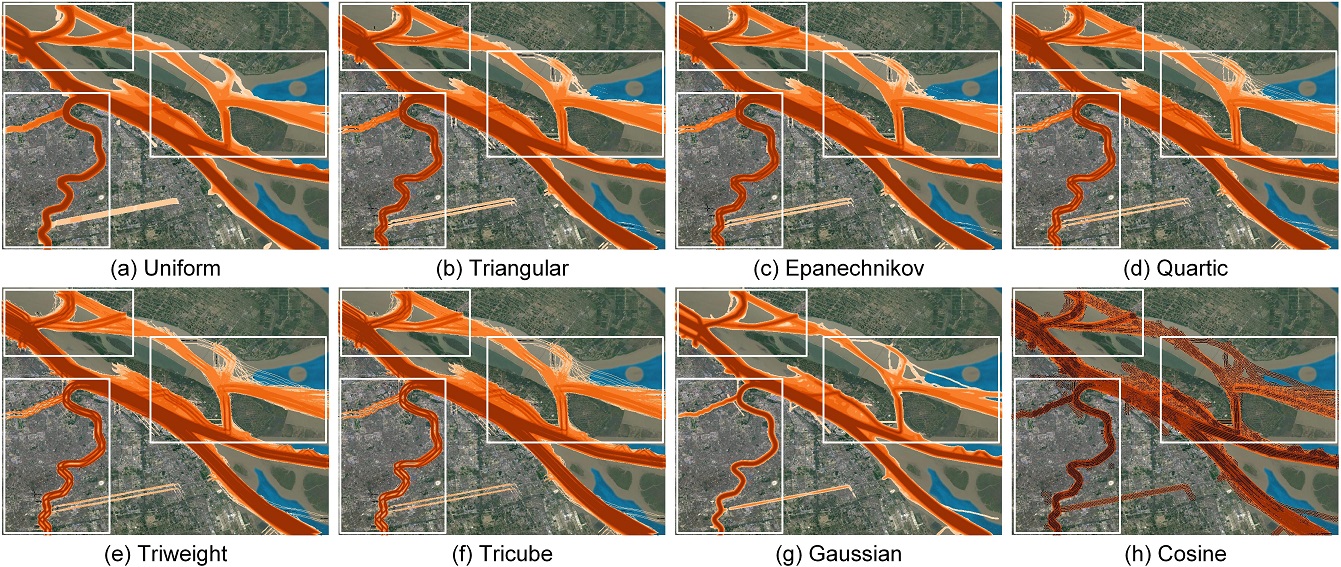}
	\caption{Visualization of large-scale vessel trajectories generated by $8$ different kernel functions of size $15 \times 15$. From top-left to bottom-right: visualization results correspond to (a) Uniform, (b) Triangular, (c) Epanechnikov, (d) Quartic, (e) Triweight, (f) Tricube, (g) Gaussian, and (h) Cosine, respectively. The white rectangles in subfigures indicate the prominent differences of visualization results.}
	\label{Fig17_Kernel15}
\end{figure*}
\begin{figure*}[!]
	\centering
	\includegraphics[width=1.00\linewidth]{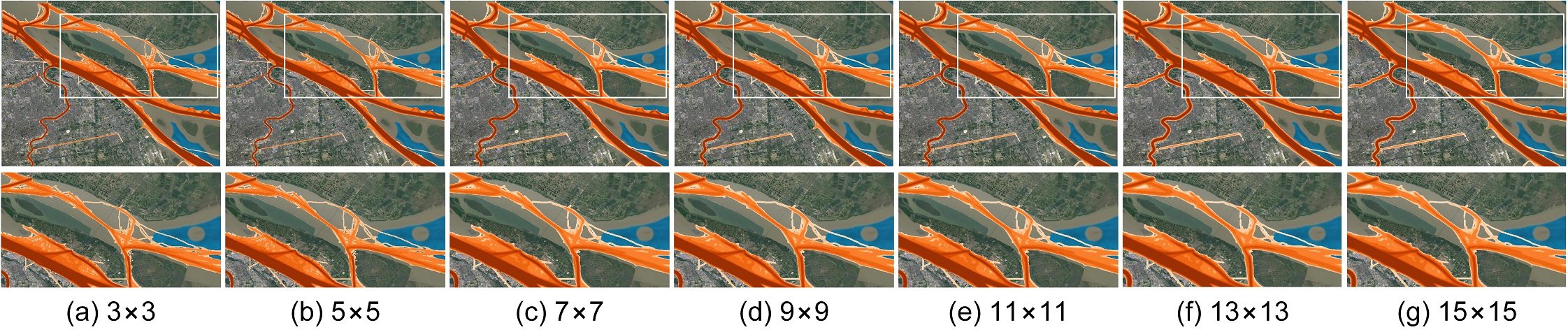}
	\caption{The comparisons of trajectory visualization results for the Gaussian kernel with $7$ different kernel sizes, i.e., (a) $3 \times 3$, (b) $5 \times 5$, (c) $7 \times 7$, (d) $9 \times 9$, (e) $11 \times 11$, (f) $13 \times 13$ and (g) $15 \times 15$. The white rectangles in subfigures indicate the prominent differences of visualization results.}
	\label{Fig18_GaussianKernel}
\end{figure*}
\subsubsection{Comparisons of Compression Costs}
To evaluate the computational robustness, each trajectory compression method runs $30$ times for $5$ different compression thresholds. In addition, the execution times for CPU serial implementations and GPU parallel implementations are detailedly illustrated in Fig. \ref{Fig11_DPtime}. In particular, the total execution time for our proposed framework includes the time to implement compression and report the results. As can be observed in Fig. \ref{Fig11_DPtime}, both CPU- and GPU-based computations can guarantee robust compression results under different thresholds. As the DP threshold increases, total execution times for both CPU and GPU computations are decreased accordingly. In contrast, the proposed GPU implementation yields significant speed improvement compared to conventional serial implementation. As shown in Fig. \ref{Fig12_DPratio}, compared with common CPU serial implementations, GPU parallel scenarios routinely yield speedups between $10.8$ and $15.3$ times for large-scale vessel trajectories in the South Channel of Yangtze River Estuary. For the Chengshan Jiao Promontory and Zhoushan Islands, our GPU-based parallelization framework can achieve the SRs between $10.1$ and $13.9$ times, and between $9.6$ and $13.7$ times, respectively. The higher speedup can be consistently achieved when the trajectories are less compressed (i.e., smaller DP threshold). The highest version is related to the smallest threshold $\epsilon = 0.1$ in this work. It can be also found that the speedup ratio of the South Channel of Yangtze River Estuary is slightly higher than that of other water areas. The reason behind this phenomenon is that the number of vessel trajectories in the South Channel of Yangtze River Estuary is larger than that in other two water areas. It means that our superiority will be more significant with the size of vessel trajectories being increased. This benefits from the massively parallel computing power of GPU for large-scale trajectory data processing.
%
% Therefore, the data transfer cost between main memory and global memory may become more significant in the overall execution time, which restricts the increase of speedup.
%
%
\subsubsection{Evaluation of Coalesced Memory Access Scheme}
In our experiments, the delay mainly refers to the memory access latency. It can be quantitatively evaluated by measuring the data transfer time between host and device memories. As shown in Fig. \ref{Fig13_OpMemory}, our GPU-based trajectory compression framework with memory access optimization could significantly reduce the transfer execution time for different compression thresholds. In contrast, the same framework without optimization suffers from the high data transfer cost.

From a theoretical point of view, the delay only contributes a small part of the total execution time. As shown in Fig. \ref{Fig14_OpTime}, it can be found that with the DP compression threshold increases, the execution time is reduced accordingly. It means that our GPU-accelerated computational framework can robustly perform trajectory compression for different thresholds. Compared with the same parallelization framework without memory access optimization, our coalesced memory access scheme is able to shorten the computational time. The difference of execution time between these two schemes is not significant since the delay constitutes only a small part of the total execution time.
% \usepackage{multirow}
%
\begin{figure*}[t]
	\centering
	\includegraphics[width=1.0\linewidth]{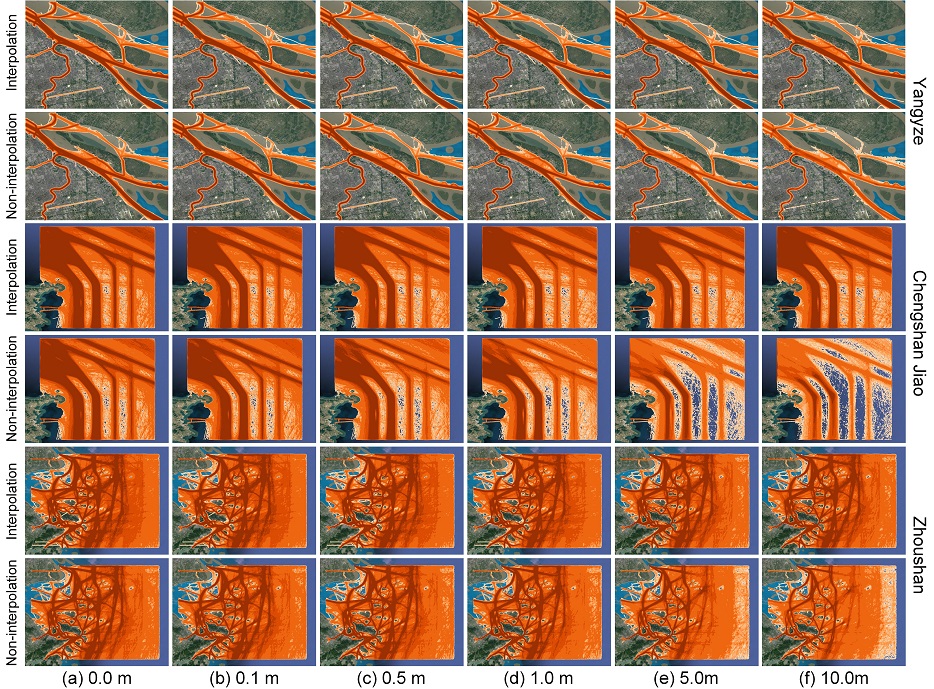}
	\caption{The comparisons of trajectory visualization results with/without interpolation operation for different compressed trajectories in $3$ different water areas, i.e., South Channel of Yangtze River Estuary (Top), Chengshan Jiao Promontory (Middle), and Zhoushan Islands (Bottom). The compressed trajectories are produced by the DP algorithm with $6$ different DP thresholds, i.e., (a) $0.0$m, (b) $0.1$m, (c) $0.5$m, (d) $1.0$m, (e) $5.0$m and (f) $10.0$m. The threshold of $0.0$m essentially indicates the original vessel trajectories without compression operation.}
	\label{Fig19_Vis3Area}
\end{figure*}
\subsection{Experiments on KDE-Based Trajectory Visualization}
\label{sec:expKDE}
\subsubsection{Influences of Kernel Functions on Visualization}
As mentioned in Section \ref{sec:KDETraj}, KDE-based vessel trajectory visualization highly depends upon the shapes of kernel functions (i.e., type and size). To determine the optimal kernel function utilized in this work, we will investigate the influences of kernel functions on visualization performance. In particular, the visualization experiments will be implemented on $8$ different kernel functions with different sizes ranged from $3 \times 3$ and $15 \times 15$. The mathematical definitions of utilized kernel functions have been illustrated in Table \ref{Table1}. In our experiments, we only select the original vessel trajectories from the South Channel of Yangtze River Estuary to investigate the influences of kernel functions on vessel trajectory visualization. Furthermore, all the water areas will be simultaneously considered to evaluate the effectiveness of the optimum kernel function in terms of visualization performance and computational cost.

The vessel density visualization results on different shapes of kernel functions are visually displayed in Figs. \ref{Fig15_Kernel3}-\ref{Fig17_Kernel15}. To avoid the negative effects of DP compression, we propose to directly generate the visualization results based on raw vessel trajectories in this subsubsection. As shown in Fig. \ref{Fig15_Kernel3}, almost no difference exists between these visualization results for the small kernel size of $3 \times 3$. With the size increases, the visualization performance generated by Cosine kernel will easily suffer from the unwanted black artifacts, illustrated in Fig. \ref{Fig16_Kernel7} (h). In contrast, the similar appearance could be found in the traffic densities visualized by other kernel functions. For the kernel size of $15 \times 15$, the black artifacts become more noticeable for Cosine kernel in Fig. \ref{Fig17_Kernel15} (h). For the sake of better comparison, the white rectangles have been adopted to highlight the regions of interest in the traffic density mappings. From Fig. \ref{Fig17_Kernel15}, it can be found that Triangular, Epanechnikov, Quartic, Triweight, Tricube and Cosine kernels unsatisfactorily cause the unnatural striping artifacts in the density mappings of vessel traffic. In contrast, both Uniform and Gaussian kernels are able to eliminate these limitations for different kernel sizes. However, the Uniform kernel, which utilizes equal weight throughout the averaging window, tends to oversmooth some structural details in final visualization results. The experiment results show that the Gaussian kernel has the capacity of robustly visualizing the vessel traffic density no matter what the kernel size is. Without loss of generality, the Gaussian kernel will be selected to implement trajectory visualization due to its stability and simplicity.

The influences of different kernel sizes on visualization are visually presented in Fig. \ref{Fig18_GaussianKernel}. In particular, the density mappings of vessel traffic are generated for $7$ different sizes ranged from $3 \times 3$ to $15 \times 15$. It can be observed that the small kernel sizes (e.g., $3 \times 3$ and $5 \times 5$ ) easily lead to the unsmooth visualization of low-density areas. As the size increases, the visualization appearance will become more natural-looking, but at the expense of improving computational cost. To achieve a proper balance between visualization quality and computational time, the Gaussian kernel with size of $7 \times 7$ will be directly utilized in this work.
\begin{figure}[t]
	\centering
	\includegraphics[width=1.00\linewidth]{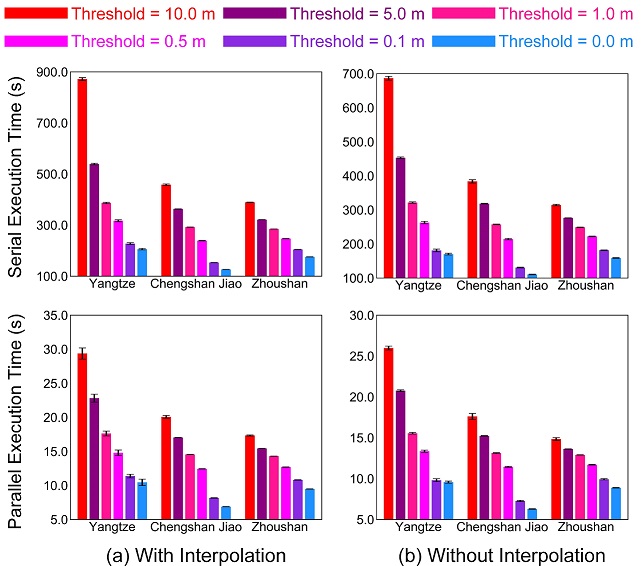}
	\caption{The comparisons of CPU and GPU execution times for visualization of compressed trajectories with/without interpolation operation in $3$ different water areas. The compressed trajectories are yielded by the DP algorithm with $6$ different DP thresholds, i.e., $0.0$m, $0.1$m, $0.5$m, $1.0$m, $5.0$m and $10.0$m.}
	\label{Fig20_KDEtime}
\end{figure}
\begin{figure}[t]
	\centering
	\includegraphics[width=1.00\linewidth]{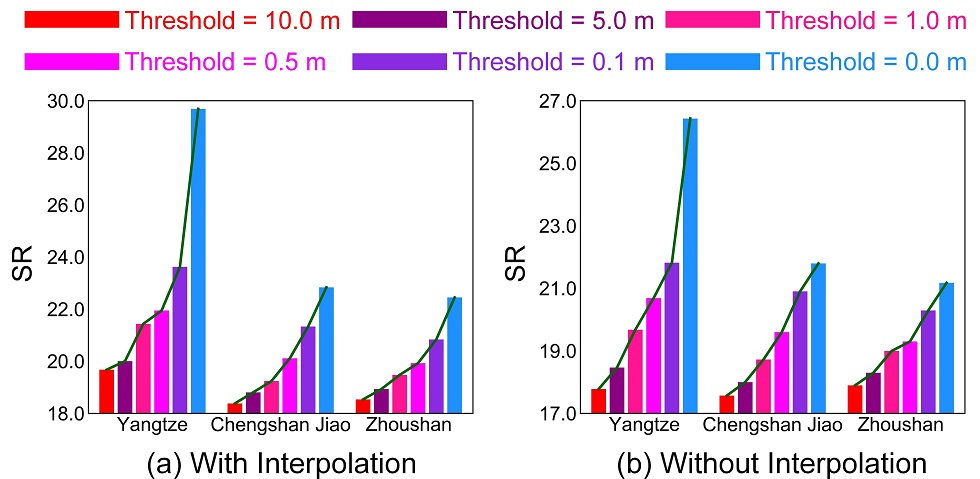}
	\caption{The speedup ratio (SR) of our GPU-based visualization framework for compressed trajectories with/without interpolation operation in $3$ different water areas. The compressed trajectories are yielded by the DP algorithm with $6$ different DP thresholds, i.e., $0.0$m, $0.1$m, $0.5$m, $1.0$m, $5.0$m and $10.0$m.}
	\label{Fig21_KDEratio}
\end{figure}
\subsubsection{Comparisons of Visualization Performance} \label{KDE Experiment}
In this experiment, we focus on investigating the influences of compression and interpolation on visualization of large-scale vessel trajectories in $3$ different water areas. According to previous experiments, the Gaussian kernel of size $7 \times 7$ is selected as the optimum kernel function. To better analyze the effects of compressed trajectories and their interpolated versions on visualization, a group of comparing experiments have been performed to visualize the vessel traffic density for $6$ different DP compression thresholds $\epsilon$ ranged from $0.0$m to $10.0$m. The visualization results implemented on compressed (i.e., no-interpolated) and interpolated vessel trajectories are displayed in Fig. \ref{Fig19_Vis3Area}.

For the original trajectories (i.e., $\epsilon = 0$), the differences of visualizations between interpolation and non-interpolation operations can be negligible for Gaussian kernel. From Fig.\ref{Fig19_Vis3Area}, it can be observed that the density mapping of vessel traffic with a DP threshold below $1.0$m is similar to the scenario yielded from original trajectories. As the threshold $\epsilon$ increases, the density mapping will not be continuous in the regions with low-density traffic. The significantly increase of RLL and DTW, shown in Table \ref{Table4}, also proves this observation. The reason is that if the threshold is excessive, the similarity between original and compressed trajectories will significantly become smaller, which further influence the density mappings of vessel traffic. It is worth noting that when the threshold $\epsilon$ is small, compressed trajectories would not lead to significantly negative effects on trajectory visualization. However, as the threshold $\epsilon$ increases, the quality of visualization on compressed trajectories will be greatly degraded. Fortunately, the introduction of interpolation operation is able to reconstruct the vessel trajectories while obviously improving the visualization performance. It means that the interpolation can reduce the negative influences of excessive compression on trajectory visualization.
\subsubsection{Comparisons of Visualization Costs}
To evaluate the visualization costs, we tend to calculate the execution times of visualization performed on compressed (i.e., no-interpolated) and interpolated vessel trajectories in $3$ different water areas. The compressed trajectories are generated through the DP algorithm with $6$ different compression thresholds $\epsilon$ ranged from $0.0$m to $10.0$m. In addition, the linear interpolation algorithm, which has been proven to be simple and effective in this work, is adopted to generate interpolated trajectories based on compressed trajectories. Each visualization experiment runs $30$ times for $6$ different DP thresholds. The execution times of visualization for both interpolated and no-interpolated are detailedly illustrated in Fig. \ref{Fig20_KDEtime}. It can be found that the interpolation operation brings a little increment in visualization costs. The speedup ratios between CPU and GPU implementations for different DP thresholds are displayed in Fig. \ref{Fig21_KDEratio}. As can be observed, the changes of speedup ratios for both interpolation and non-interpolation operations have the similar regularity. It means that the interpolation operation significantly enhances the visualization quality but leading to negligible additional computational cost. In summary, DP compression is able to significantly accelerate the visualization of large-scale vessel trajectories in maritime IoT industries. The additional visualization cost yielded by interpolation operation could be negligible in practical applications.
%
% Note that when the threshold decreases from $0.1$m to $0.0$m, the speedup ratio increases significantly. The compression ratio of $59.25\%$ shown in Table \ref{Table4} could confirm this observation \textcolor[rgb]{1,0,0}{in the South Channel of Yangtze River Estuary}. 
%
%In summary, the experimental results not only prove the superior acceleration performance of proposed KDE-based parallelization algorithm, but also illustrate that DP-based compression under threshold within a certain range do reduce the computing complexity of KDE-based visualization while ensuring the data quality.
%
%\section{Conclusion and Future Work}
\section{Conclusion}
Compression and visualization of vessel trajectories have become significantly important in handling data mining problems in maritime IoT industries. However, large-scale vessel trajectories commonly lead to high computational cost, which make efficient implementations of compression and visualization challenging tasks. To tremendously shorten the execution time, we proposed to develop the GPU-based parallelization frameworks by taking full advantages of the massively parallel computation capabilities of GPU architecture. In particular, the proposed frameworks could dramatically accelerate DP and KDE algorithms for trajectory compression and visualization, respectively. Numerous experiments have been performed to demonstrate the effectiveness in terms of reducing execution time and guaranteeing compression and visualization results. With the size of vessel trajectories becomes larger, our superiority will be more significant in the era of big data.% To further enhance our proposed methods, the studies presented in this work could be extended along the following directions:
\label{section5}
\ifCLASSOPTIONcaptionsoff
  \newpage
\fi
\end{document}